\documentclass[11pt,preprint]{aastex}
\usepackage{times}
\usepackage{natbib}
\usepackage{amsmath}

\newcommand{\str}{Str\"{o}mgren }
\newcommand{\ars}{\langle R_s \rangle}
\newcommand{\nseven}{n_{\rm H, \infty} = 10^7$~cm$^{-3}}
\newcommand{\nsix}{n_{\rm H, \infty} = 10^6$~cm$^{-3}}
\newcommand{\nfive}{n_{\rm H, \infty} = 10^5$~cm$^{-3}}
\newcommand{\lambdaavg}{\langle \lambda_{\rm rad} \rangle}
\newcommand{\lambdamax}{\lambda_{\rm rad, max} }
\newcommand{\hi}{\rm H~{\textsc i}}
\newcommand{\hii}{\rm H~{\textsc {ii}}}
\newcommand{\mtwo}{M_{\rm bh,2}}

\shorttitle{ACCRETION ONTO BLACK HOLES REGULATED BY RADIATIVE FEEDBACK}
\shortauthors{Park \& Ricotti}

\received{}
\accepted{}
\begin{document}

\title{ACCRETION ONTO BLACK HOLES FROM LARGE SCALES REGULATED BY RADIATIVE FEEDBACK. II. \\ GROWTH RATE AND DUTY CYCLE }

\author{KwangHo Park and Massimo Ricotti} 
\affil{Department of Astronomy, University of Maryland, College Park, MD 20740, USA} 
\email{kpark@astro.umd.edu, ricotti@astro.umd.edu}

\begin{abstract}
This paper, the second of a series on radiation-regulated accretion onto
black holes (BHs) from galactic scales, focuses on the effects of radiation
pressure and angular momentum of the accreting gas. We simulate
accretion onto intermediate-mass black holes, but we derive general
scaling relationships that are solutions of the Bondi problem with
radiation feedback valid for any mass of the BH $M_{\rm bh}$.
Thermal pressure of the ionized sphere around the BH regulates
the accretion rate producing periodic and short-lived luminosity bursts.
We find that for ambient gas densities exceeding $n^{\rm cr}_{\rm
H,\infty} \propto M_{\rm bh}^{-1}$, the period of the oscillations
decreases rapidly and the duty cycle increases from 6\%, in
agreement with observations of the fraction of active galactic nuclei
at $z\sim 3$, to 50\%. The mean accretion rate becomes Eddington limited for
$n_{\rm H,\infty}>n_{\rm H,\infty}^{\rm Edd} \simeq n_{\rm H,\infty}^{\rm
cr} T_{\infty,4}^{-1}$ where $T_{\infty,4}$ is the gas temperature
in units of $10^4$~K. In the sub-Eddington regime, the mean accretion rate
onto BHs is about $1\% T_{\infty,4}^{2.5}$ of the Bondi rate,
thus is proportional to the thermal pressure of the ambient medium. The
period of the oscillations coincides with depletion timescale of the
gas inside the ionized bubble surrounding the BH. Gas depletion
is dominated by a pressure gradient pushing the gas outward if $n_{\rm
H,\infty}<n^{\rm cr}_{\rm H,\infty}$ and by accretion onto the BH 
otherwise.  Generally, for $n_{\rm H,\infty}<n^{\rm cr}_{\rm H,\infty}$
angular momentum does not affect significantly the accretion rate and
period of the oscillations.

\end{abstract}


\section{INTRODUCTION} 
Gravitational accretion onto point sources can be described
analytically \citep{BondiH:44,Bondi:52} assuming spherical symmetry.
The Eddington-limited Bondi formula is often prescribed in
cosmological simulations to estimate the gas accretion rate onto black
holes (BHs) from large scales
\citep*{Volonteri:05,DiMatteo:08,Pelupessy:07,Greif:08,AlvarezWA:09,KimWAA:11}.
However, even in sub-Eddington regime, a fraction of the gravitational
potential energy of accreted gas is converted into mechanical or
radiative feedback \citep{Shapiro:73}, reducing the accretion rate.  The
radiation emitted by BHs creates feedback loops that regulate the gas
accretion and luminosity of the BHs. Several published works have
investigated physical processes that may dominate the feedback such as
X-ray preheating, gas cooling, photo-heating, and radiation pressures
\citep*{OstrikerWYM:76,CowieOS:78,BB:80,KrolikL:83,Vitello:84,WandelYM:84,
  MiloBCO:09,OstrikerCCNP:10,NovakOC:11}. In general, radiative
feedback reduces the accretion luminosity of the accreting BH
\citep*{OstrikerWYM:76,Begelman:85,Ricotti:08}. There have been
extensive publications on self-regulation of supermassive BH 
growth at the centers of elliptical galaxies
\citep*{Sazonov:05,CiottiO:07,CiottiOP:09,LussoC:11} or axisymmetric outflows in
active galactic nuclei
\citep*[AGNs;][]{Proga:07,ProgaOK:08,KurosawaPN:09,KurosawaP:09a,KurosawaP:09b}.
Recently, several works have paid closer attention to
radiation-regulated accretion onto intermediate-mass black holes
\citep*[IMBHs;][]{MiloBCO:09,MiloCB:09,ParkR:11,Li:11}.

Cosmological simulations show that massive BHs may have formed in
metal-free minihalos as Population III star remnants in the early universe
\citep*{AbelANZ:98,BrommCL:99,AbelBN:00,MadauR:01,SchneiderFNO:02,OhH:02}
or from direct collapse of primordial gas
\citep*{Carr:84,HaehneltNR:98,Fryer:01,BegelmanVR:06,VolonteriLN:08,
  OmukaiSH:08,ReganH:09,MayerKEC:10,JohnsonKGD:11}. Estimating the
accretion luminosity of IMBHs \citep[for a review, see][]{Miller:04,vanderMarel:04} 
is important to understand their
cosmological importance at high $z$ and in the local universe
\citep*{MackOR:07,Ricotti:09}. Since the luminosity of IMBHs is
directly related to their accretion rate, these studies are also
relevant for better understanding the mass growth of primordial
massive BHs in the early universe
\citep*{MadauR:01,VolonteriHM:03,YooM:04,Volonteri:05,JohnsonB:07,Pelupessy:07,
AlvarezWA:09} or provide clues about the origin and impact on the
ionization history of the universe of ultraluminous X-ray sources
\citep*[ULXs;][]{Krolik:81,Krolik:84,Krolik:04,RicottiO:04b,
RicottiOG:05,Ricotti:07,StrohmayerM:09}.

In \citet*{ParkR:11}, hereafter Paper~I, we explored the
radiation-regulated accretion onto IMBHs assuming spherical symmetry
and zero angular momentum of the accreting gas. One of the main
objectives of the study was to derive an analytical description of the
effect of radiation feedback on the Bondi accretion rate. We
accomplished this goal by simulating accretion onto IMBHs with
idealized initial conditions and simple physics, but for a large
parameter space of the initial conditions (varying BH radiative
efficiency, the BH mass, the density and temperature of ambient gas,
and the spectrum of radiation). We found that the IMBH is quiescent
most of the time with short intense periodic bursts of accretion (with duty
cycle of 6\%). The qualitative description of the cycle is as
follows. Gas accumulates in a dense shell ahead of the I-front
nearly halting accretion onto the IMBH. Meanwhile, the ionized gas inside
the hot bubble is pushed outward toward the dense shell by pressure
gradients, eventually de-pressurizing the hot bubble, producing the
collapse of the shell and a burst of accretion. The scaling relationships for
the burst period, mean and peak accretion rates can be understood
analytically but are quite sensitive to the details of the thermal
structure inside the \str sphere. Thus, we expect that gas
metallicity may be an important parameter in the problem that 
we have not yet explored.  

In this paper, the second of the series, we relax most of the
simplifying assumptions in Paper~I and discuss the effects of helium
heating/cooling, radiation pressure, and gas angular momentum on the
accretion rate.  It has been noted that not only electron scattering
but also radiation pressure on \hi~may be important
\citep{MiloBCO:09}. We explore how the radiation pressure regulates
the gas accretion by transferring momentum to the inflowing gas, and
whether these physical processes become important compared to the
pressure gradients inside the \str sphere. 

As in Paper~I of this series, here we present the results of our
simulations as dimensionless accretion rates $\lambda_{\rm rad} \equiv
\dot{M}/\dot{M}_B$, where $\dot{M}_B = \pi e^{3/2} \rho_\infty
G^2M_{\rm bh}^2 c_{s,\infty}^{-3}$ is the Bondi accretion rate for an
isothermal gas (with polytropic index $\gamma = 1$), that is a
function of BH mass $M_{\rm bh}$, density $\rho_{\infty}$, and sound
speed of $c_{s,\infty}$ of neighboring gas. The Eddington luminosity
is $L_{\rm Edd}= 4 \pi G M_{\rm bh} m_p c \sigma_{\rm T}^{-1}$
, which is proportional to the BH mass only. 
We define the Eddington accretion rate $\dot{M}_{\rm
Edd} \equiv L_{\rm Edd}/c^2$ which is a factor of 10 smaller than the
definition most often used $\dot{M}_{\rm Edd} \equiv L_{\rm Edd}/0.1 c^2$,
in which $\eta=0.1$ is assumed, and 
the dimensionless Eddington rate as $\lambda_{\rm Edd}
\equiv \eta^{-1}\dot{M}_{\rm Edd}/\dot{M}_{B}$.  We introduced three
quantities to describe the accretion: mean accretion rate
$\lambdaavg$, accretion rate at peaks $\lambdamax$, and period between
accretion bursts $\tau_{\rm cycle}$.  In Paper~I we found 
\begin{equation}
\lambdaavg \simeq C(n_{\rm H,\infty})~T_{\rm \infty,4}^{2.5} \left[ 
T_{\rm in}(\alpha, Z)/4\times10^4~{\rm K}\right]^{-4},
\end{equation}
with $C \sim 3\%$ for $n_{\rm H,\infty}\ge 10^5$~cm$^{-3}$ and $C \sim 3\%
(n_{\rm H,\infty}/10^5~{\rm cm}^{-3})^{1/2}$ for 
$n_{\rm H,\infty} < 10^5$~cm$^{-3}$ with $\eta=0.1$ and 
$M_{\rm bh} = 100~M_{\odot}$. Here, $T_{\rm in}(\alpha, Z)$
is the time-averaged temperature at the accretion radius within the
\str sphere, that is $4\times 10^4$~K for our fiducial spectrum
(spectral slope $\alpha =1.5$) and a gas composed of hydrogen only
(metallicity $Z=0$). However, $T_{\rm in}$ is sensitive to changes of
$\alpha$ and the gas composition (see Paper~I and Section~3 in this
paper). In the sub-Eddington regime, the period $\tau_{\rm cycle}
\propto \ars $ where $\ars$ is the time-averaged \str radius, and
the duty cycle $f_{\rm duty} \equiv \tau_{\rm on}/\tau_{\rm cycle} \equiv
\lambdaavg/\lambdamax \sim 6\%~T_{\infty,4}^{1/2} $, where $\tau_{\rm on}$ is the duration
of bursts.

This paper is organized as follows.  In Section~2, we briefly explain
the numerical methods. In Section~3, we present our simulation results
including the effect of helium cooling/heating, radiation pressure, and
gas angular momentum. Summary and discussion are presented in
Section~4.

\begin{figure}[t]
\epsscale{1.0} \plotone{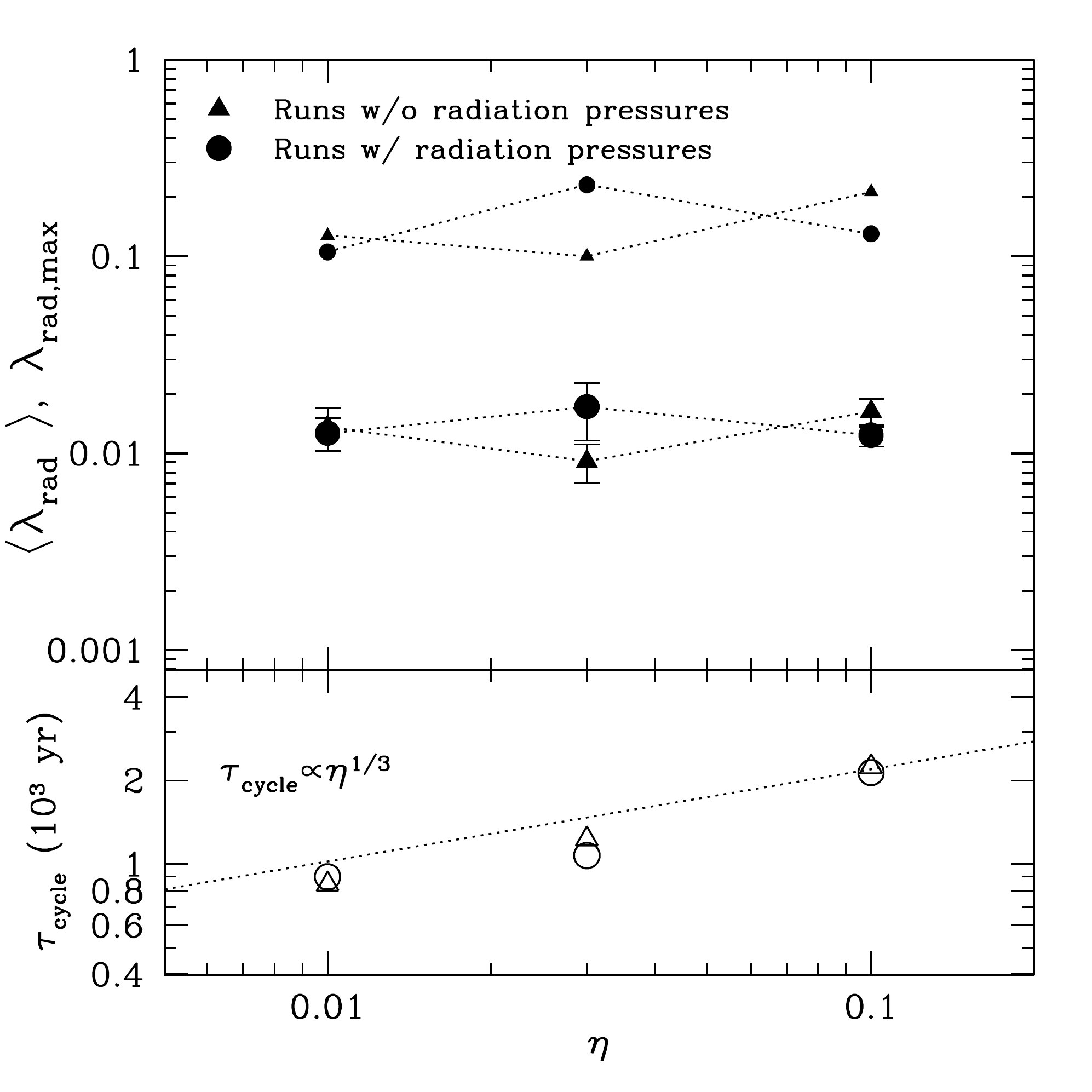}
\caption{ Top: accretion rate in units of the Bondi rate as a function of
radiative efficiency $\eta$ for simulations with $M_{\rm bh}=100~M_\odot$,
$\nsix$, and $T_\infty=10^4~$K. Large symbols indicate mean accretion rate
($\lambdaavg \sim 1\% $), while small symbols show accretion rate at peaks
 ($\lambdamax \sim 20\%$). Bottom: period between bursts $\tau_{\rm
 cycle}$ as a function $\eta$. The dotted line shows $\tau_{\rm cycle}
 \propto
\eta^{1/3}$. In both panels, triangles represent simulations neglecting
the effect of radiation pressures, while circles show simulations including
radiation pressures. Radiation pressures introduce a minor difference
in both the accretion rate and period of the bursts
for this parameter set.}
\label{eta}
\end{figure}

\section{NUMERICAL SIMULATIONS}
We run a suite of hydrodynamic simulations to interpret how radiative
feedback regulates accretion onto BHs. We use a modified parallel
non-relativistic hydrodynamics code, ZEUS-MP \citep{StoneN:92,Hayes:06} plus
a radiative transfer algorithm \citep{RicottiGS:01} 
to simulate photo-ionization and photo-heating by UV and X-ray ionizing photons
emitted near the BHs. See Paper I for a detailed description.

In this study, we include the effects of helium heating/cooing in addition
to hydrogen. Therefore, we simulate photo-ionization,
photo-heating and cooling for six species \hi, \hii, He~{\sc i},
He~{\sc ii}, He~{\sc iii}, and $e^-$.

We also calculate the radiation pressures both on $e^-$ and \hi~to
interpret the effect of momentum transfer to the inflowing gas by the
ionizing photons. The magnitude of acceleration at a given radius due
to radiation pressure depends on the luminosity, the ionization fraction
of hydrogen and helium, and the cross section of the species
to photon-ionization. The specific flux $F_\nu \propto e^{-\tau}/r^2$ 
at a given radius($r$), assuming a power-law spectrum
with a spectral index $\alpha$, depends on the optical depth  
$\tau_\nu$, and the cross section
$\sigma_\nu$. Thus, the accelerations due to momentum transfer 
to \hi~ and $e^-$ can be written as
\begin{eqnarray}
a_{\rm rad, \hi} = \frac{x_{\rm \hi}}{m_p c} \int \sigma_{\rm \hi,\nu} F_{\nu} d\nu, \\
a_{\rm rad,e^-} = \frac{x_{e^-}}{m_p c} \int \sigma_T F_{\nu} d\nu, 
\end{eqnarray}
where $x_{\rm \hi}$ and $x_{e^-}$ are \hi~and $e^-$ fractions,
respectively, $\sigma_T$ is the Thomson cross section, and $m_{p}$ is
the proton mass.  The radial component of the acceleration at a given
radius is updated as ${\bf a}={\bf a}_{\rm grav}+{\bf a}_{\rm rad}$,
where ${\bf a}_{\rm rad}= {\bf a}_{\rm rad,\hi}+{\bf a}_{\rm rad,e^-}$.

In Section~\ref{ssec:ang}, we study the effect of non-zero angular
momentum of gas which leads to a time delay between the accretion rate
at the sonic radius and the luminosity output, due to the formation of
an accretion disk. In order to estimate realistic values of the time
delay we assume that the gas conserves angular momentum and settles
into an accretion disk of radius $R_{\rm disk}$. We then assume an
alpha model for the thin disk to estimate the timescale for the gas
to lose angular momentum and fall into the BH.

Numerically, it is convenient to express the time delay in units of
the free-fall timescale $t_{\rm ff}$ calculated at the simulation's
inner boundary (typically $R_{\rm min} \sim 10^{-5}~$pc). The
free-fall timescale we have defined can be very large compared to
$t_{\rm ff}$ calculated at the radius of the accretion disk near the
BH (at tens of gravitational radii $R_{\rm Sch} \equiv 2GM_{\rm bh}/c^2$). 
Approximately, 
the gas is accreted at the viscous timescale $t_{\rm visc}$, that 
compared to $t_{\rm ff}$ is
\begin{equation}
t_{\rm visc}(R_{\rm disk})/t_{\rm ff}(R_{\rm disk}) \sim \alpha^{-1} \mathcal{M}^2 
\sim \alpha^{-1} c_{\rm s, disk}^{-2} GM_{\rm bh} R_{\rm disk}^{-1}
\sim 0.5~\alpha^{-1} (c/c_{\rm s,disk})^2 \mathcal{R}_{\rm disk}^{-1}, 
\end{equation} 
where $\alpha$ is the dimensionless parameter for a thin disk
\citep{ShakuraS:1973}, $c_{\rm s, disk}$ is the sound speed
of the gas in the disk, and we define $\mathcal{R}_{\rm disk} 
\equiv R_{\rm disk}/R_{\rm Sch}$. 
The dependence of the free-fall time on
radius is $t_{\rm ff} \propto R^{1.5}$, while the viscous timescales
as $t_{\rm visc} \propto R^{-1} t_{\rm ff} \propto R^{0.5} $ assuming
constant sound speed due to effective cooling (note that since we 
are considering a gas of zero
or very low metallicity, the gas in the disk will not easily cool to temperature
below $10^4$~K if the gas is atomic). Thus, the infall time at the disk radius 
$R_{\rm disk}$ is 
\begin{equation}
\frac{t_{\rm visc}(R_{\rm disk})}{t_{\rm ff}(R_{\rm min})}
\sim \frac{0.5}{\alpha} \frac{v_{\rm min}^3}{c~c_{\rm s,disk}^2} 
\mathcal{R_{\rm disk}}^{1/2} 
\sim \frac{0.3}{\alpha} \left(T_{\rm disk} \over 10^4~{\rm K} \right)^{-1}
\mathcal{R}_{\rm disk}^{1/2} .
\label{eq:tvisc}
\end{equation}
To estimate the parameters in Equation~(\ref{eq:tvisc}) we have defined
$v_{\rm min} \equiv (GM_{\rm bh}/R_{\rm min})^{1/2} \simeq 260~{\rm
km~s^{-1}}$. Assuming $\alpha \sim 0.01$--$0.1$, $T_{\rm disk} \sim 10^4~{\rm
K}$, and $\mathcal{R}_{\rm disk} \lesssim 10^2$--$10^4$, 
we find time delays of $\lesssim 300$
free-fall times at $R_{\rm min}$,
that is the parameter space we explore in Section~\ref{sec:tdelay}.


\begin{figure}[t]
\epsscale{1.0} \plotone{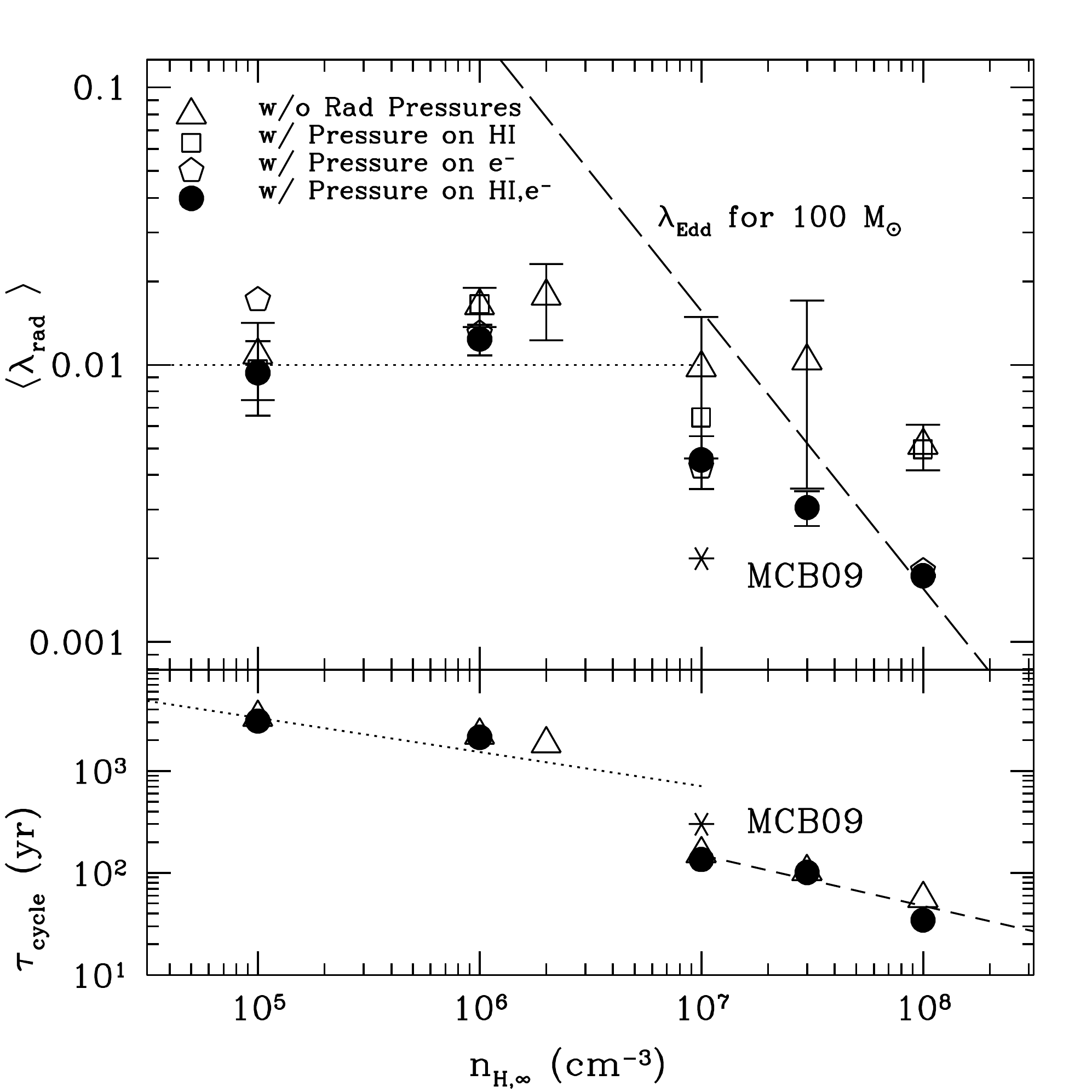}
\caption{ Comparison of relative importance of radiation pressures in
regulating mean accretion rate (top panel) and the period of bursts
(bottom panel) as a function of the ambient gas density $n_{\rm
H,\infty}$. Symbols are explained in the legend of the figure. 
Top: long dashed line 
shows the Eddington limit for a 100~$M_\odot$ BH with $\eta=0.1$.
When the accretion rate is sub-Eddington ( $n_{\rm H,\infty}
\le 10^6~{\rm cm}^{-3}$ ) radiation pressures both on electrons 
and \hi~do not play an important role and the thermal structure of 
the \str sphere regulates the accretion. Radiation pressure is 
important in reducing the accretion rate at 
$n_{\rm H,\infty} = 10^7$~cm$^{-3}$ where the accretion rate
approaches the Eddington rate. The transition of accretion rate from
$\lambdaavg \sim 1\%$ to the Eddington-limited regime happens 
at $n_{\rm H,\infty}^{\rm Edd} \sim 4\times 10^6$~cm$^{-3}$ for a 
100~$M_\odot$ BH with $\eta=0.1$ and $T_\infty=10^4$~K. Bottom: radiation
pressures do not produce significant differences in $\tau_{\rm cycle}$. 
Transition of $\tau_{\rm cycle}$ from mode-I (dotted line) 
to mode-II (short dashed line) happens at the critical density 
$n_{\rm H, \infty}^{\rm cr} \sim n_{\rm H, \infty}^{\rm Edd}$ (see
Section~\ref{sec:2regimes}). The result shows a good agreement with the work 
of \citet{MiloCB:09} }
\label{den}\end{figure}

\begin{figure}[t] 
\epsscale{1.0} \plotone{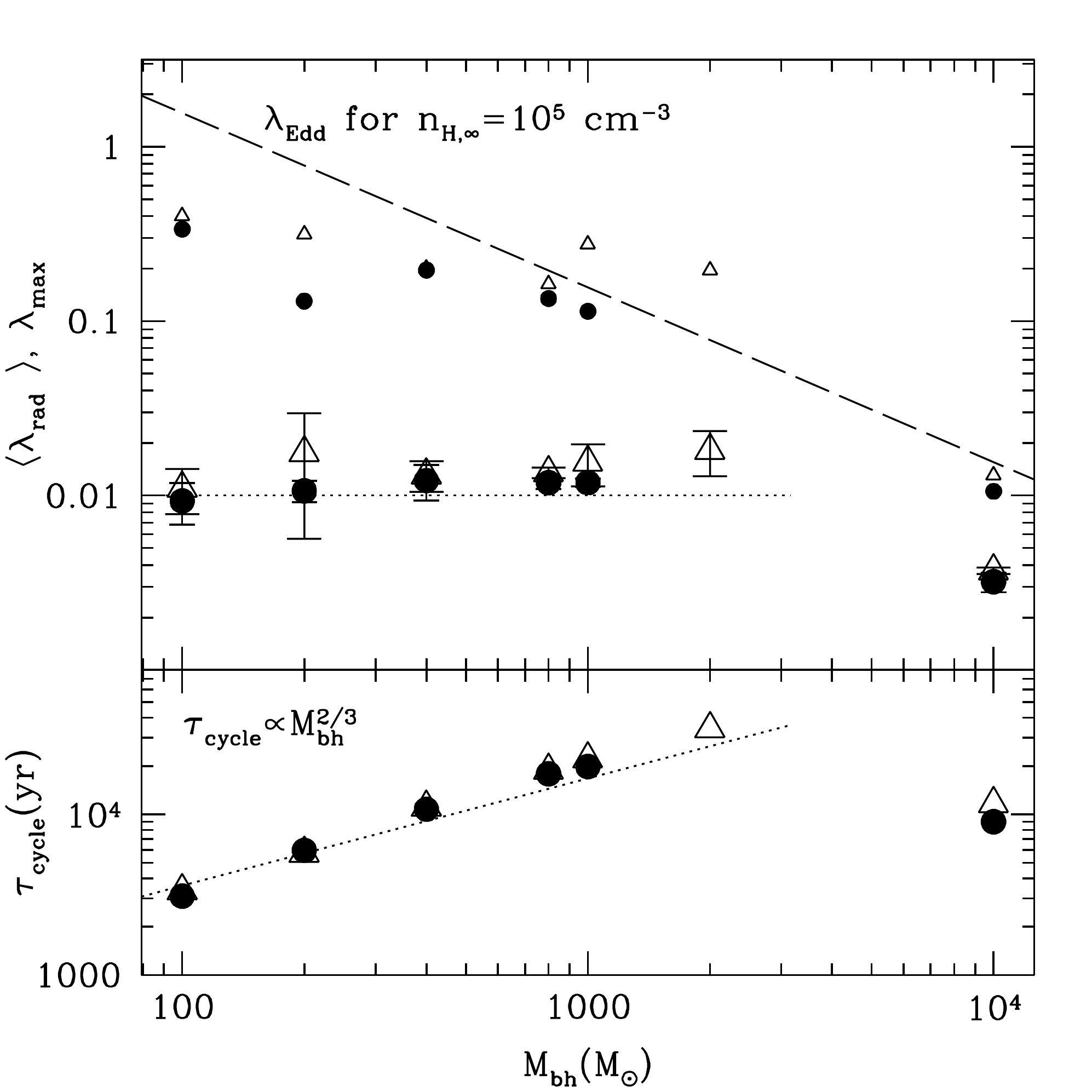} 
\caption{Same as Figure~\ref{eta}, but showing $\lambdaavg$, $\lambdamax$, and 
$\tau_{\rm cycle}$ as a function of $M_{\rm bh}$ with $\eta=0.1$, $\nfive$,
and $T_{\infty}=10^4$~K. A similar pattern which we observe as a function
of density is also seen as a function of $M_{\rm bh}$. With increasing
$M_{\rm bh}$, the transition from $\lambdaavg \sim 1\%$ to the 
Eddington-limited regime and the transition of $\tau_{\rm cycle}$ from
mode-I to mode-II happen at $M_{\rm bh} \sim 4\times 10^3~M_\odot$.}
\label{mass} \end{figure}

In our code, the accretion
rates calculated at the inner boundary of the simulations are stored in
an assigned array about 1000 steps for each $t_{\rm
  ff}$. Stored accretion rates with a given time delay are then read
from the array and used to estimate the luminosity at the current
moment.

\begin{figure*}[t]
\epsscale{1.0} \plotone{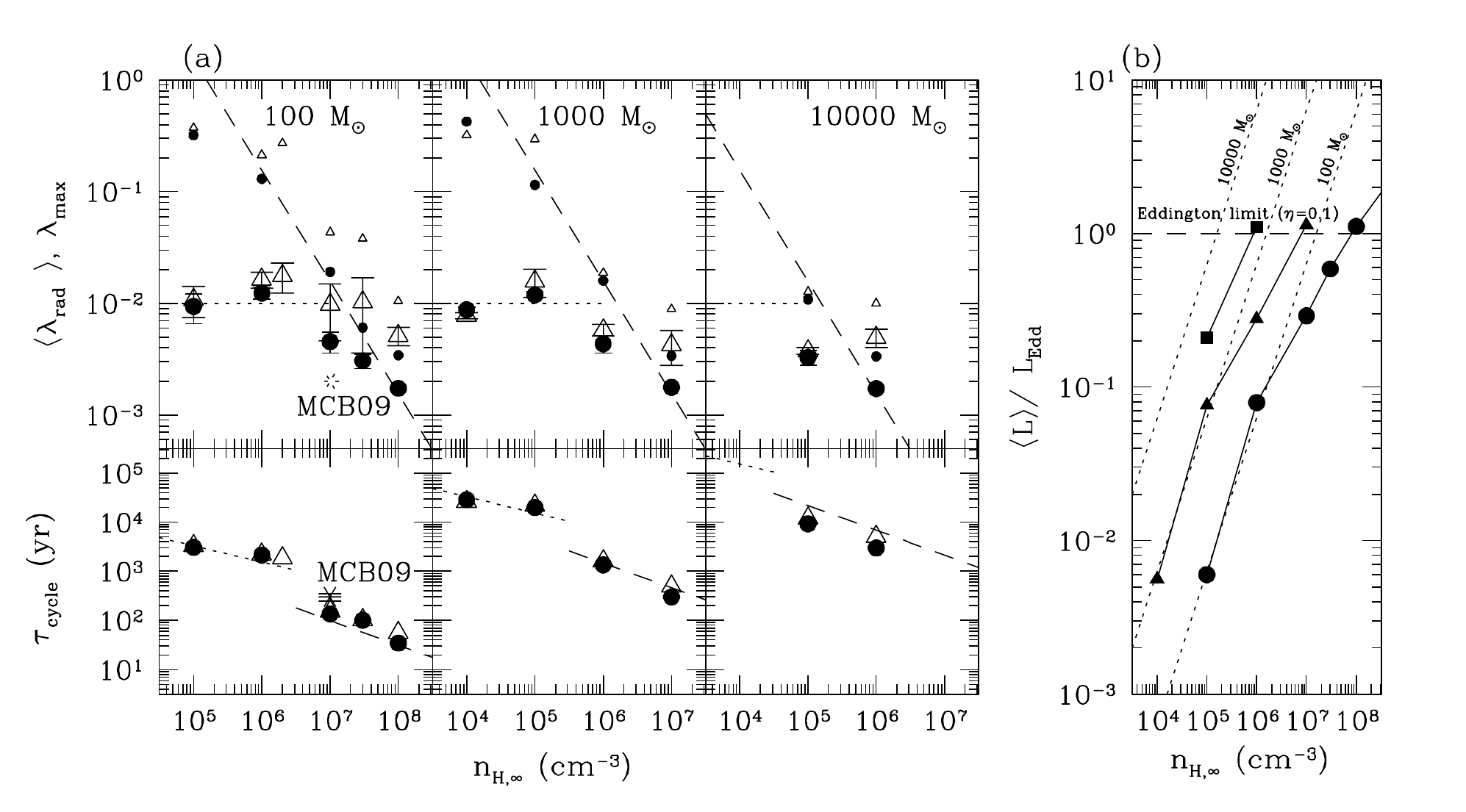}
\caption{ Left: same as Figure~\ref{den}, but showing $\lambdaavg$,
$\lambdamax$, and  $\tau_{\rm cycle}$ as a function of gas
density for $M_{\rm bh}=10^2~M_\odot$ (left panel), $10^3~M_\odot$ 
(middle panel), and $10^4~M_\odot$ (right panel). 
Long dashed lines in each panel show the Eddington
accretion rate for $\eta=0.1$ and the given BH mass. 
With increasing gas density, the accretion rate eventually becomes 
Eddington limited, but the transition to the Eddington-limited regime occurs
at densities $n_{\rm H,\infty}^{\rm Edd} \sim 4\times 10^6~{\rm cm}^3 
M_{\rm bh}^{-1}$ decreasing linearly with increasing BH mass.
Right: accretion luminosities normalized by Eddington luminosities for the same
simulations in the left figure. 
Symbols (circles: $10^2~M_\odot$, triangles: $10^3~M_\odot$,
squares: $10^4~M_\odot$) show the simulations including radiation pressures
for each BH mass. With increasing gas density, the accretion rate
becomes regulated primarily by Compton radiation pressure.} 
\label{den_mass} \end{figure*}

\section{RESULTS}

In this section, we show the results of simulations discussing the
effects of helium heating/cooling, radiation pressures, and angular
momentum on the BH accretion rate. All the simulations in this paper include
helium heating/cooling but the gas is metal free. The simulations
remain qualitatively the same as in Paper~I where we did not include
helium; the only noticeable difference with respect to Paper~I is that
the accretion rate at peak luminosity shows multiple minor peaks
instead of a well-defined single peak. 
This is to be expected, as a larger opacity produces a stronger feedback
with respect to the hydrogen only case, leading to multiple shocks in
the gas. This complicate structure -- i.e., a burst consisting of several
sub-bursts -- is commonly found \cite[e.g.,][]{CiottiO:07}.
In addition, the average
accretion rate $\lambdaavg$ decreases from $\sim 3\%$ to $\sim 1\%$,
but this can be understood by the increase of the mean temperature
inside the \str sphere to from $T_{\rm in} \sim 4 \times 10^4$~K to 
$\sim 6 \times 10^4$~K. The top panel
of Figure~\ref{eta} shows the accretion rate as a function of
$\eta=0.01$-$0.1$ with $M_{\rm bh}= 100~M_\odot$, $\nsix$, and
$T_{\infty}=10^4~{\rm K}$.  Large symbols show $\lambdaavg$ while
small symbols show $\lambdamax$.  For the given set of parameters, the
luminosity remains in the sub-Eddington regime, thus the effects of
radiation pressures are minor. The bottom panel of Figure~\ref{eta}
shows the dependence of $\tau_{\rm cycle}$ on $\eta^{1/3}$, the same
as found in Paper~I. However, $\tau_{\rm cycle}$ for $\eta =0.1$ is
now $\sim 2200$~years which is $\sim 60\%$ of the value found in
Paper~I for the given set of parameters. This is also well understood
(see Equation~(22) in Paper~I) as our model predicts $\tau_{\rm cycle}
\propto \lambdaavg^{1/3}$.

\subsection{Effect of Radiation Pressures}
In Paper I, we have focused on exploring the parameter space in which
the mean accretion rate is dominated by thermal feedback, i.e.,
radiation pressure can be neglected. We found $\lambdaavg \sim 1\%$
for $n_{\rm H, \infty} = 10^5$~cm$^{-3}$, assuming
$M_{\rm bh}=100~M_\odot$, $T_\infty=10^4$~K, $\alpha=1.5$, 
and including helium cooling/heating.
However, not surprisingly, including the effect of radiation
pressure produces a reduction of the accretion rate when the BH luminosity
approaches the Eddington limit.
Figure~\ref{den} shows $\lambdaavg$ as a function of gas
density for a $100~M_\odot$ BH, comparing simulations that do not
include radiation pressure (open triangles) to ones including pressure
on \hi~only (open squares), on $e^-$ only (open pentagons), and the
total effect of radiation pressure (solid circles). Compton radiation
pressure reduces the accretion rate below $\lambdaavg \sim 1\%$ for
$n_{\rm H, \infty} \gtrsim 10^7$~cm$^{-3}$ while the radiation
pressure on \hi~ appears always negligible with respect to Compton
scattering. Both $\lambdaavg$ and $\lambdamax$ change from a
constant fraction of the Bondi accretion rate to the Eddington rate
$\lambda_{\rm Edd}$, shown by the dashed line for
$M_{\rm bh}=100~M_\odot$ and radiative efficiency $\eta = 0.1$.

Figure~\ref{mass} shows the dimensionless accretion rates $\lambdaavg$
and $\lambdamax$ as a function of the BH mass from $M_{\rm
  bh}=10^2$ to $10^4~M_\odot$, keeping the other parameters constant:
$\eta=0.1$, $\nfive$, and $T_{\infty}=10^4$~K. The simulations include
radiation pressures on \hi~and $e^-$, and show that the transition
to Eddington-limited accretion happens for $M_{\rm bh} \simeq
5000~M_\odot$.

\subsubsection{Transition from Bondi-like to Eddington-limited Accretion}

So far the simulation results have shown that Compton scattering on
electrons is the dominant radiation pressure effect, thus the
Eddington-limit applies. Figure~\ref{den_mass} summarizes the results of
a large set of simulations that include radiation pressure. The top
three panels in Figure~\ref{den_mass}(a) shows $\lambdaavg$ as a
function of gas density for $M_{\rm bh} =10^2, 10^3,{\rm
  and}~10^4~M_{\odot}$, respectively. For each BH mass, corresponding
Eddington limits are shown by the dashed lines. The panels show the
mean accretion rate $\lambdaavg$ (large triangles) and $\lambdamax$
(small triangles) transitioning from being a constant fraction of the
Bondi rate at low densities to being Eddington-limited at higher
densities. The period of the accretion $\tau_{\rm cycle}$, in the bottom
panels, also shows different dependencies in Bondi-like and
Eddington-limited regimes. We will come back to this in Section~\ref{sec:2regimes}.

Figure~\ref{den_mass}(b) shows the mean accretion luminosity in
units of $L_{\rm Edd}$ for $M_{\rm bh} =10^2, 10^3,{\rm
  and}~10^4~M_{\odot}$ as a function of gas density. The dotted lines
show $1\%$ of the Bondi accretion rate for each BH mass.  Thus, from
Figure~\ref{den_mass} approximately we have
\begin{equation}
\langle \dot{M} \rangle = {\rm min}(1\% T_{\rm \infty,4}^{2.5} \dot{M}_{B}, \eta^{-1} \dot{M}_{\rm Edd}),   
\end{equation}
where $T_{\infty,4} \equiv T_{\infty}/(10^4~{\rm K})$, valid for
density $n_{\rm H, \infty} \gtrsim 10^5~{\rm cm^{-3}}$, and
$\alpha=1.5$.  

It is thus apparent that IMBHs can grow at a rate near the Eddington
limit if the gas density of the environment is larger than the
critical density
\begin{equation} 
n_{\rm H,\infty}^{\rm Edd} \sim 4 \times 10^6~{\rm cm}^{-3} \left(
\frac{M_{\rm bh}}{10^2~M_\odot}\right)^{-1} \left(
\frac{T_\infty}{10^4~{\rm K}}\right)^{-1} \left(
\frac{\eta}{0.1}\right)^{-1}.
\label{eq:bondi_edd} 
\end{equation}


\subsubsection{Why is Continuum Radiation Pressure Negligible?}
\begin{figure*}[t]
\epsscale{1.0} \plotone{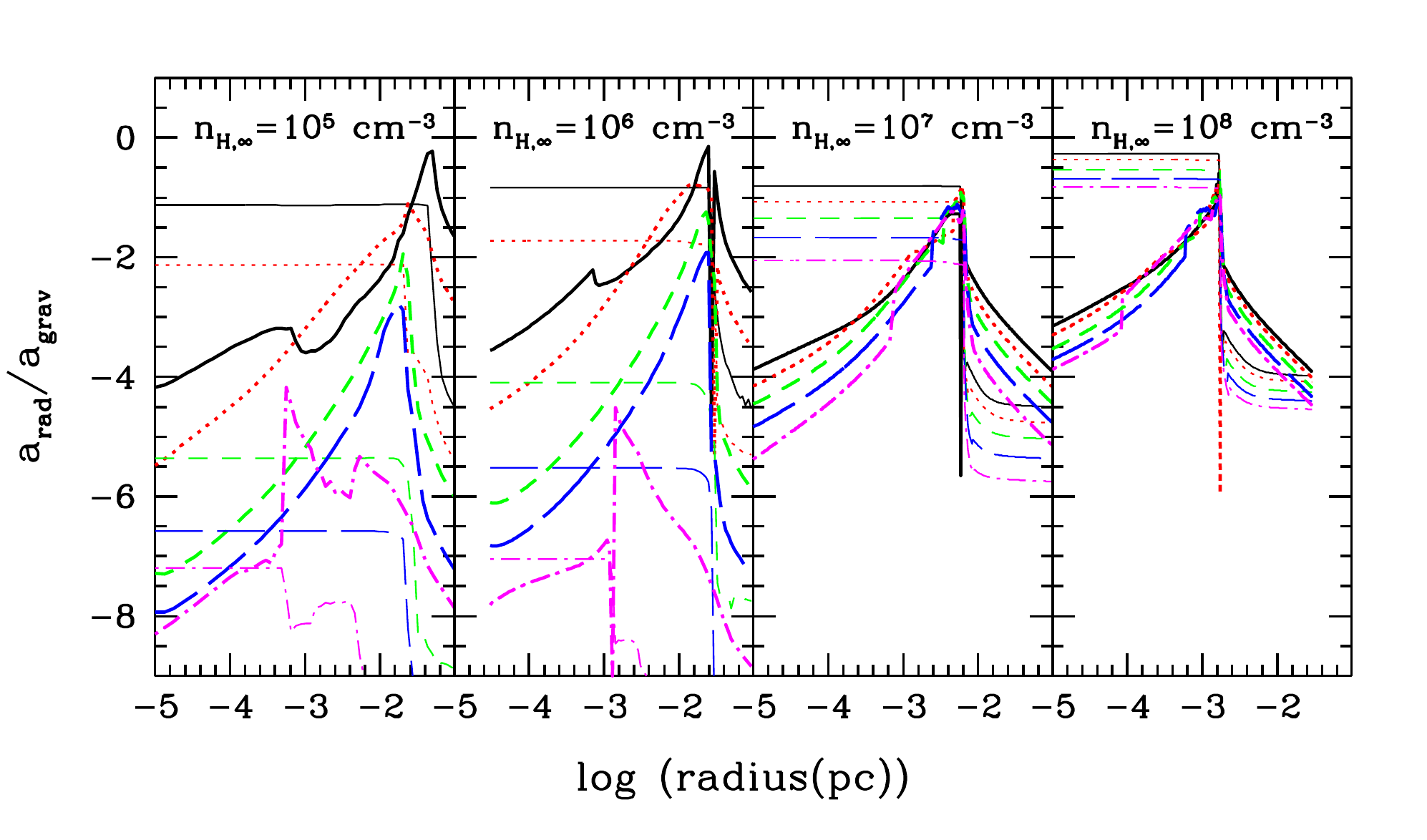}
\caption{ Radial profiles of the gas acceleration due to radiation
pressures on \hi~ and $e^-$ normalized to the gravitational acceleration
of simulations for $n_{\rm H,\infty}=10^5, 10^6, 10^7, {\rm and}~10^8~{\rm
cm^{-3}}$ with BH mass $M_{\rm bh}=100~M_\odot$, $\eta=0.1$, and
$T_{\infty}=10^4$~K. Thick lines refer to radiation pressure on \hi,
while thin lines show Compton scattering radiation pressure. Different line
types show the profiles
at different time during the oscillation cycle 
(e.g., solid lines at the accretion
bursts and dot-dashed just before the bursts). 
Radiation pressure on \hi~ peaks just
inside the \str sphere with weak dependence on density. While Compton 
radiation pressure inside the \str sphere increases on average as a function
of density. At $\nseven$ the peak values of \hi~radiation
pressure and Compton radiation pressure become comparable and about 10\%
of gravity. However, at higher densities ($\ge 10^8$~cm$^{-3}$) 
radiation pressure on electrons becomes dominant everywhere inside 
the \str sphere.}
\label{rp1} \end{figure*}

As shown in Figure~\ref{den}-\ref{den_mass}, the
simulations show that radiation pressure on \hi~does not play an
important role when the accretion rate is sub-Eddington. In this
section, we focus on understanding why this is.  Figure~\ref{rp1} shows
the evolution of relative magnitude of acceleration due to radiation
pressures normalized by the gravitational acceleration at a given
radius. Each panel refers to a different density $n_{\rm
  H,\infty}=10^5, 10^6, 10^7, {\rm and}~10^8~{\rm cm^{-3}}$. Within
the \str sphere, the relative effect of Compton radiation pressure
remains constant as a function of the radius since the electron fraction
$x_{e^-}$ is close to unity and the gas is nearly transparent to
ionizing radiation. Outside of the \str sphere, the rapid decrease of
the electron fraction reduces the effect of Compton scattering.
Radiation pressure on \hi~(thick lines in Figure~\ref{rp1}) increases
as a function of radius and has its peak value just inside the \str
sphere. This is due to the increase of the \hi~ fraction as a function
of radius. Outside the \str sphere the relative effect of \hi~
radiation pressure drops quickly because the ionizing luminosity
decreases rapidly due to the increase of the \hi~ opacity.

Continuum radiation pressure on \hi~is comparable to Compton electron
scattering only in a shell just inside the \str sphere, where the
\hi~abundance starts to increase rapidly as a function of radius and
the ionizing radiation is not fully shielded by \hi. With increasing
gas density, the peak and mean luminosities increase, hence the
relative effect of Compton pressure on average increases and
eventually becomes comparable to the effect of gravity (i.e., Eddington
limit). Whereas Figure~\ref{rp1} shows that the relative effect of
continuum radiation pressure does not increase much with increasing
gas density. In addition, the range of variation of radiation
pressures during a period of oscillation decreases with increasing
density. In other words, at low densities radiation pressures display
several magnitudes of variation which are not seen in the high-density
regime. As a result, at low densities ($n_{\rm H, \infty} \lesssim
10^6~{\rm cm}^{-3}$) radiation pressure is significant only near the
peaks of luminosity and generally is negligible compared to gravity;
whereas at high densities ($n_{\rm H, \infty} \gtrsim 10^7~{\rm
  cm}^{-3}$) Compton scattering dominates throughout a period of
oscillation reducing the accretion rate to Eddington-limited values.
Only at intermediate densities $n_{\rm H, \infty} \simeq 10^7~{\rm cm}^{-3}$, 
the magnitude of \hi~radiation pressure
just behind the \str radius becomes comparable to that by Compton
scattering.

The weak dependence of the \hi~radiation pressure on density and its
magnitude with respect to the Compton pressure can be understood
analytically. The key point is that the \hi~radiation pressure is
proportional to the value of the neutral fraction $x_{\rm \hi}$ just
behind the \str radius $R_s$ and, assuming ionization equilibrium, it
is easy to show that $x_{\rm \hi}(R_s) \propto n_{\rm H}^{-2/3}$.
It follows that the
pressure on \hi~is relatively insensitive to variations of $n_{\rm H}$:
\begin{equation}
P^{\rm cont}_{\rm Rad} \propto S_0 x_{\rm \hi}(R_s) \exp{[-\tau(R_s)]} 
\propto n_{\rm H}^{1/3},
\end{equation}
where $S_0 \propto n_{\rm H}$ is the ionizing luminosity, and
$\exp{[-\tau(R_s)]}={\rm const}$.  The derivation of $x_{\rm \hi}(R_s)$ is as
follows. At $R_s = S_0^{1/3} n_{\rm H}^{-2/3} \alpha_R^{-1/3}$ the
photoionization rate is $\Gamma(R_s) = S_0 \sigma^{\rm eff}_{\rm
  \hi}/4\pi R_s^2 \propto n_{\rm H}^{5/3}$. Assuming photoionization
equilibrium $x_{\rm \hi}(R_s)\Gamma(R_s) = n_{\rm H} \alpha_R$, we demonstrate
that
\begin{equation} 
x_{\rm \hi} (R_s) = \frac{n_{\rm H}\alpha_R}{\Gamma(R_s)} \propto n_{\rm H}^{-2/3}.
\end{equation}

\subsection{Two Self-regulated Modes of Accretion: Collapsing I-front versus Quasi-steady I-front}\label{sec:2regimes} 

One of the most interesting aspects of the radiation-regulated
accretion onto BHs is the qualitative change of the period and duty
cycle of the luminosity bursts observed in the high-density regime. As
argued in Paper~I and confirmed by further simulations in this work,
the physical reason for this transition is a change of the dominant
mechanism depleting the gas inside the \str sphere between two
consecutive bursts. In the low-density regime, gas is pushed outward
toward the ionization front by a pressure gradient (hereafter, mode-I
accretion). At higher-densities gas accretion onto the BH becomes the
dominant gas depletion mechanism (hereafter, mode-II
accretion). Incidentally, as discussed below, simulations show that
radiation pressure becomes important near the transition from mode-I
to mode-II, at least for most of the initial conditions we have simulated.
\begin{figure*}[t]
\epsscale{1.0} \plotone{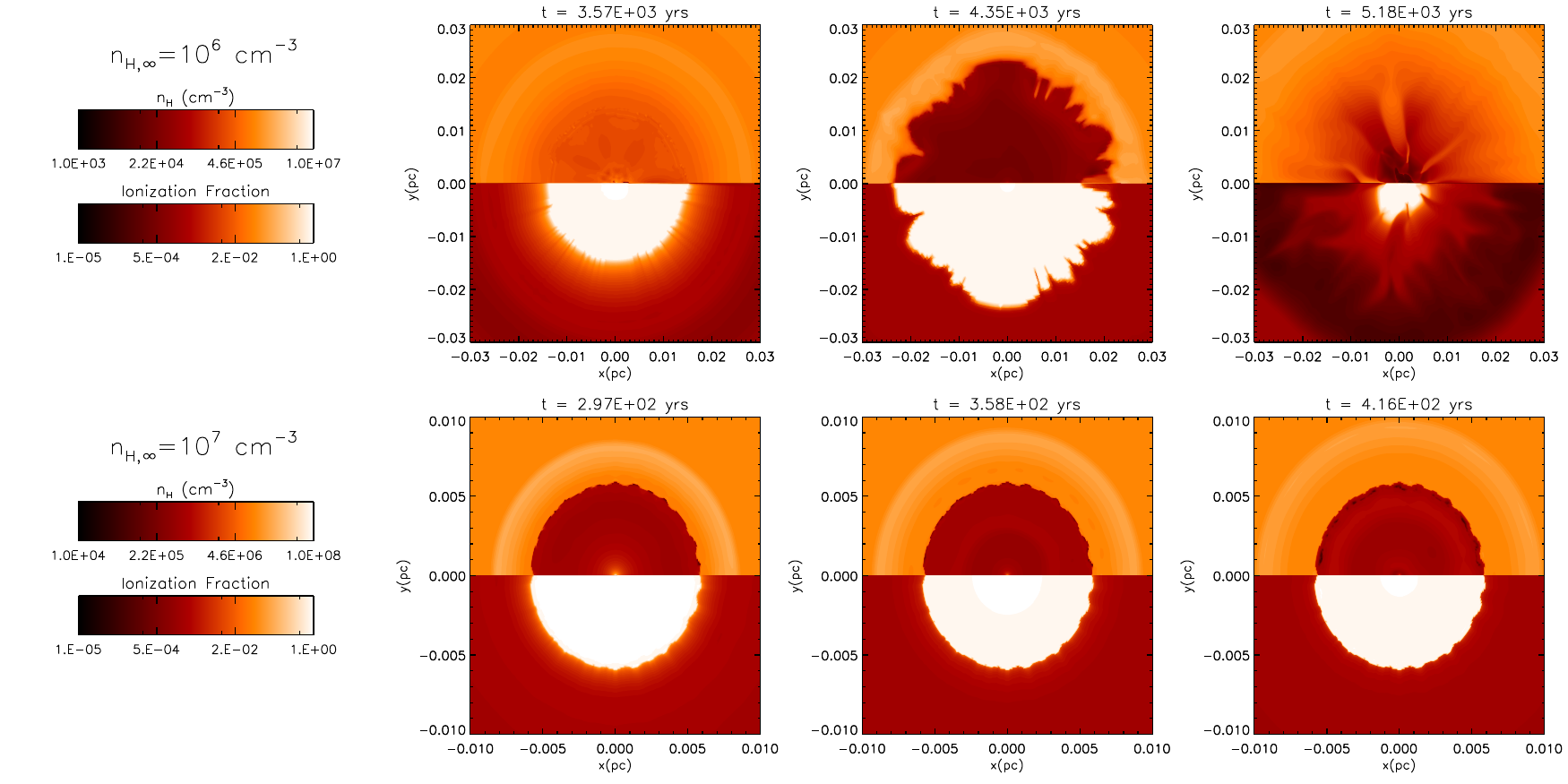}
\caption{Illustration of the two different modes of oscillations found
at different ambient densities. The panels show the time evolution of
gas density and ionization fraction in 2D simulation for a BH of mass
$M_{\rm bh}=100~M_\odot$, gas density $\nsix$ (top panels), and $\nseven$
(bottom panels).  In each panel, top halves show the density and the
bottom halves show the ionization fraction, 
$x_{\rm \hii} = n_{\rm \hii}/n_{\rm H}$. When
the density is $n_{\rm H, \infty} \le n_{\rm H, \infty} ^{\rm cr} \simeq
5\times 10^6~{\rm cm^3}$, the collapse of ionization front onto the BH
leads to a burst of accretion luminosity (mode-I). For densities $n_{\rm
H, \infty} > n_{\rm H, \infty}^{\rm cr} $ the size of \str sphere does
not change much with time (mode-II). Note the different oscillation
modes of the accretion rate and luminosity are driven by the collapse 
of dense shell (mode-I) and by a density wave (mode-II).}
\label{evolution} \end{figure*}
In Paper~I, we have observed mode-II accretion only for our highest density
simulation (for $\nseven$ and $M_{\rm bh}=100~ M_\odot$).  In
this paper, to better understand this regime, we have extended the
parameter space to higher densities and higher BH masses.
Figure~\ref{evolution} shows snapshots of the density (top halves in
each panel) and ionization fraction (bottom halves in each panel) for
two-dimensional simulations including radiation pressure, 
for $\nsix$ (top panels)
and $\nseven$ (bottom panels). The snapshots are taken for each
simulation at the moment of a burst of the accretion rate (left
panels), in-between two bursts (middle panels), and just before a
burst (right panels). For ambient density $\nsix$, the \str sphere
collapses onto the BH which leads to a strong luminosity burst. On the
contrary, the size of \str sphere does not change much during the
oscillation period for ambient density $\nseven$. In this latter case,
the oscillation of the accretion luminosity is driven by density and
pressure waves originating at the I-front, while in the former case,
the collapse of the I-front onto the BH leads to a much more intense
accretion burst. In Figure~\ref{d6d7}(left), we compare the accretion
rate onto the BH as a function of time for $\nsix$ (top panel) and
$\nseven$ (bottom panel). For $\nsix$, the collapse of I-front leads
to strong burst of gas accretion, with $\lambdamax$ about $\times
20 \lambdaavg$. Hence, the duty cycle $f_{\rm duty}^{\rm I} \equiv
\lambdaavg/\lambdamax$ is about $6\%$. The pressure gradient inside
\str sphere supports the gas shell accumulating at the I-front from
collapsing until the accretion rate drops 4-5 orders of
magnitude compared to the accretion during the burst. However, the
\str radius remains remarkably constant before its collapse due to the
decline of gas density inside the \hii~region. In contrast, in the
$\nseven$ simulation the accretion rate peaks at a few times
$\lambdaavg$ before decreasing by about 2 orders of magnitude. The
duty cycle approaches $f_{\rm duty}^{\rm II} \sim 50\%$ for this mode of accretion. As shown in
Figures~\ref{mass} and \ref{den_mass}, simulations that do not
include radiation pressure also show a rapid decrease of the period
$\tau_{\rm cycle}$ and $\lambdamax$ with increasing gas density and
BH mass, but the mean accretion rate $\lambdaavg$ does not. Thus, the
reduced value of $\lambdamax/\lambdaavg \equiv 1/f_{\rm duty}^{\rm II}$ 
explains the longer duty cycle observed for mode-II accretion.
A more detailed illustration of the qualitative difference between
mode-I and mode-II accretion is shown in Figure~\ref{d6d7}(right). The
figure shows the time evolution of the gas density profile (top
panels), the temperature profile (middle panels) and the hydrogen
ionization fraction (bottom panels) for the $\nsix$ and $\nseven$
simulations. Small variations of the density, temperature, and
ionization fraction profiles are observed for $\nseven$, while clear
collapses of I-front are observed in the evolution of the profiles for
$\nsix$. Note that this quasi-stationary profile is not produced by
the effects of radiation pressures. The same effect is found for
$\nseven$ without including radiation pressure effects.

Interestingly, for our fiducial case simulations ($M_{\rm bh}=100~M_\odot$,
$T_{\infty}=10^4$~K, $\eta=0.1$, and $\alpha=1.5$), the critical
density at which the mean accretion rate becomes Eddington-limited
nearly coincides with the critical density for transition from mode-I
to mode-II accretion. This explains why the mean accretion rate and
the peak accretion rate become Eddington-limited at nearly the same
density. Indeed, if while increasing $n_{\rm H,\infty}$, 
the duty cycle remained at about $6\%$ as
in mode-I accretion, the mean accretion rate would not be able to
approach the Eddington limit, even though the peak accretion can be
mildly super-Eddington. We will show below that the transition to
mode-II accretion depends on the free parameters in the problem and
may take place at much lower densities than the critical density for
Eddington-limited accretion.

The quasi-stationary I-front observed for the $\nseven$ simulation is
also important to understand why there exists a clear transition to
the Eddington-limited regime with increasing density or BH mass. For
mode-I accretion, radiation pressure may become comparable to the
gravity near the \str radius, but this effect dominates only for a
short time, during the peaks of luminosity. The peak accretion can
indeed become moderately super-Eddington for a short time, also
because of the broken spherical symmetry of the
collapsing shell due to Rayleigh--Taylor instability of the accreting
gas. However, for
mode-II accretion, the geometry of accretion from large scales is
quasi-spherical and radiation pressure effects are significant during
the most of the duration of oscillations, hence the accretion rate is
Eddington limited.
\begin{figure*}[t]
\epsscale{1.0} \plottwo{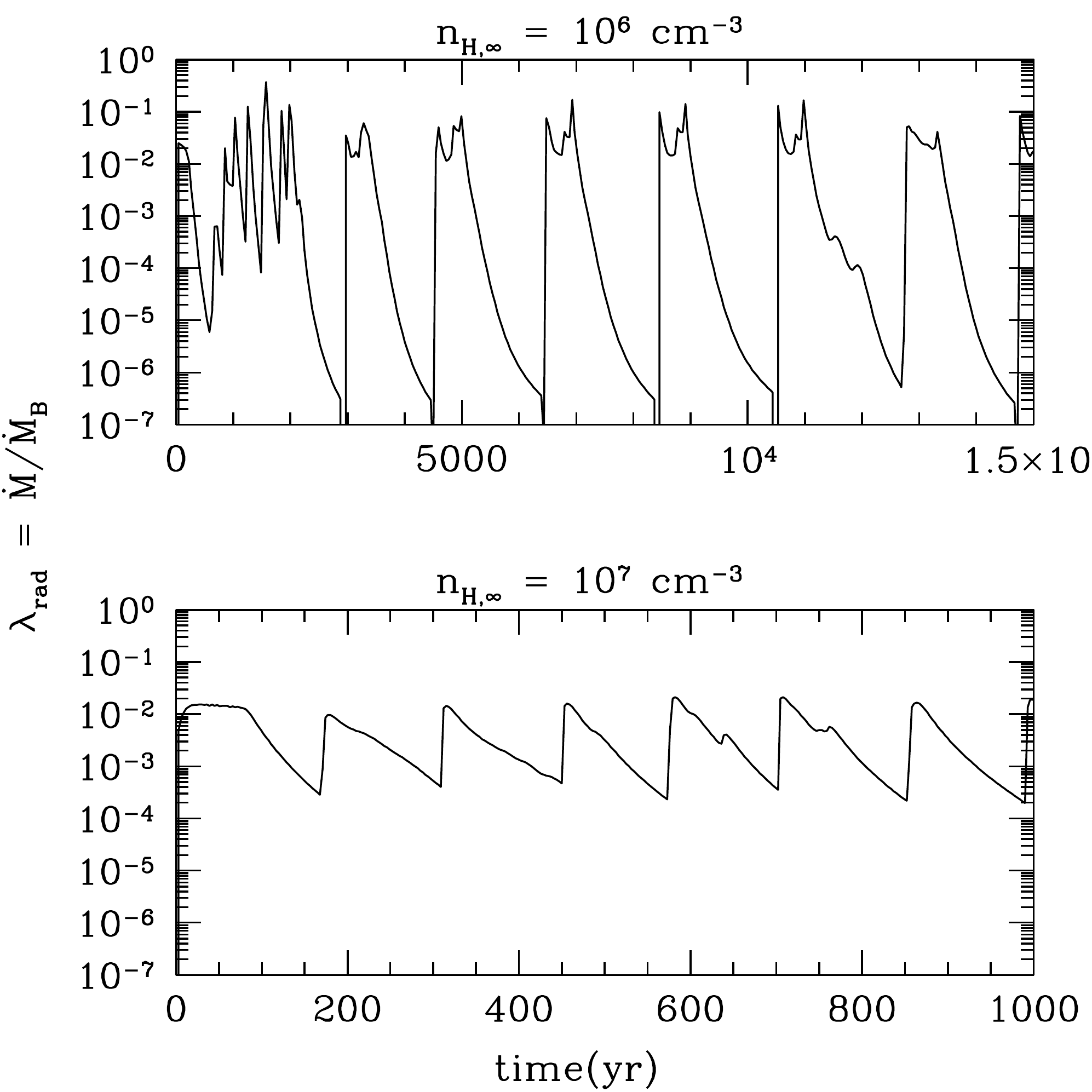}{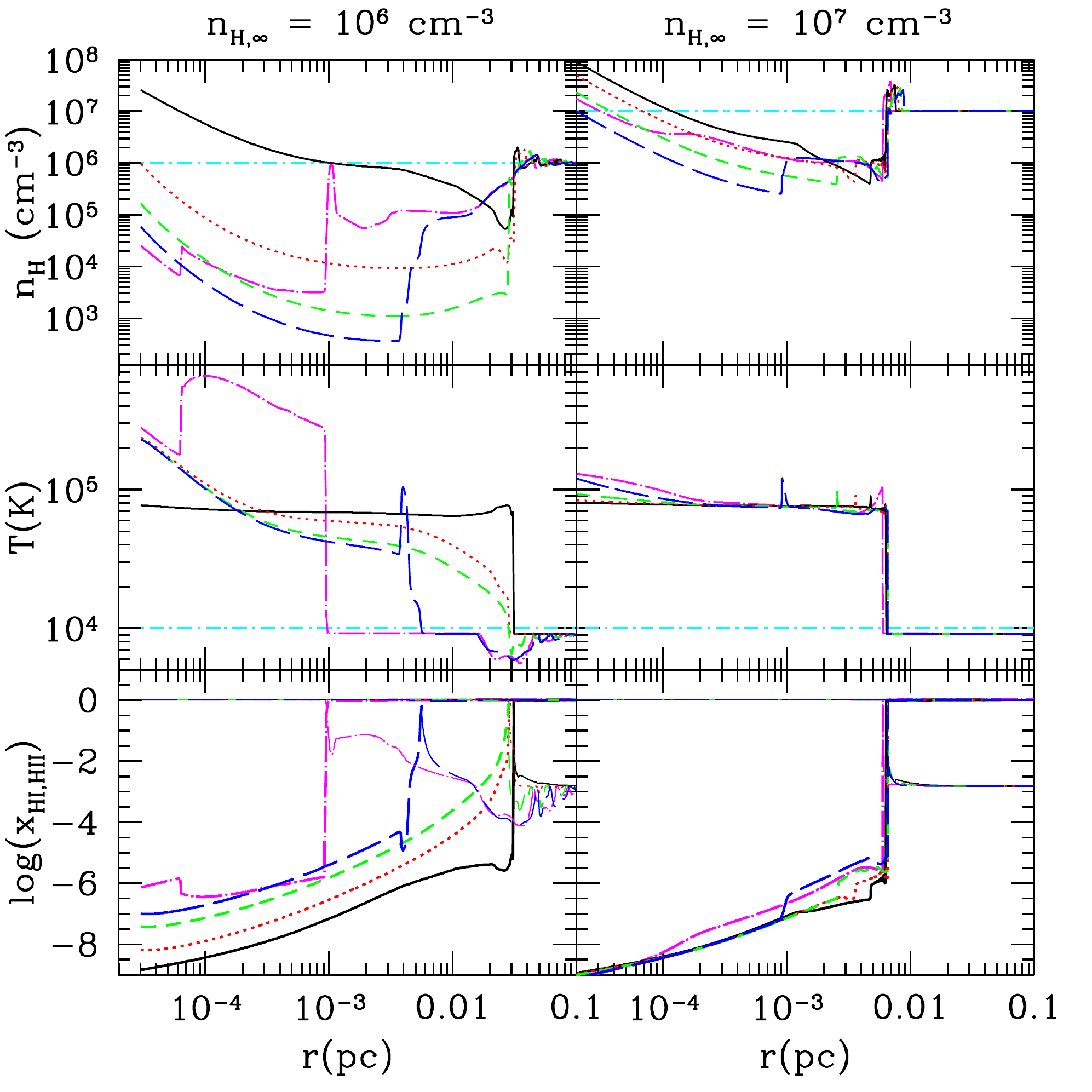}
\caption{ Left: accretion rates as a function of time for $\nsix$ and
$\nseven$ assuming $\eta=0.1$, $M_{\rm bh}=100~M_\odot$, and 
$T_\infty=10^4~{\rm K}$. 
Different modes of oscillations occur at different density
regimes. Mode-I oscillation at $\nsix$ shows about 5 orders of magnitude
range between peak and the minimum accretion rate, while mode-II 
oscillation at $\nseven$ shows only 2 orders of magnitude range. 
Right: evolution of radial profiles for density (top panel), 
temperature (middle panel), and
neutral/ionization fractions (bottom panel) of the same simulations in the left
figure. Note the change
of physical properties inside \str sphere during a period of mode-I 
oscillation ($\nsix$), while mild changes are observed for mode-II
oscillation ($\nseven$). }
\label{d6d7} \end{figure*}
\begin{figure}[t]
\epsscale{1.0} \plotone{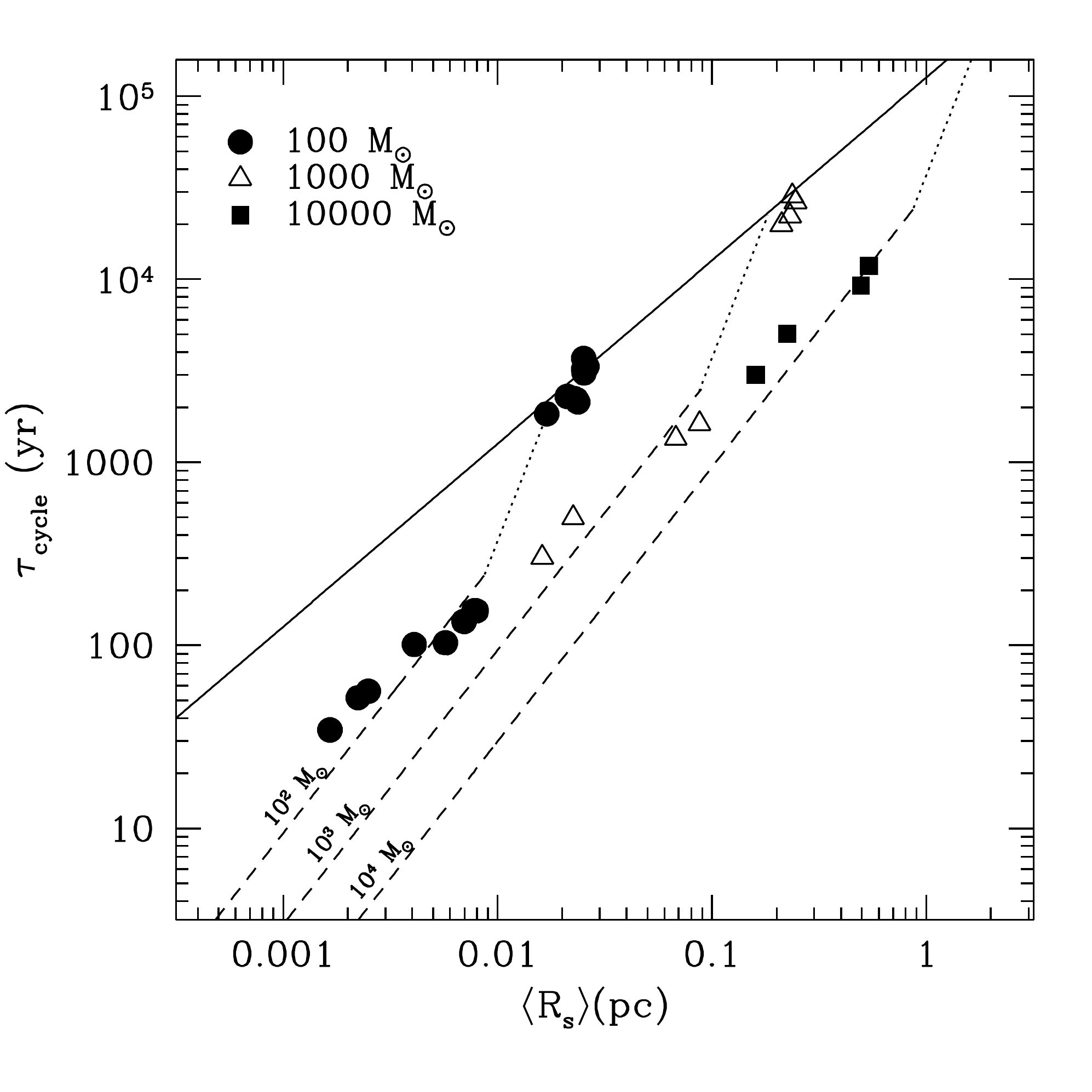}
\caption{Relationship between the period of the accretion bursts, $\tau_{\rm
cycle}$, and the time-averaged size of \str radius $\ars$. $\tau_{\rm
cycle}$ shows a linear relation with $\ars$ when the gas depletion 
inside the \str sphere is dominated by a pressure gradient inside the
\hii~region that push the gas toward the I-front. Instead,   
$\tau_{\rm cycle} \propto \ars^3$ (dotted lines for each $M_{\rm bh}$) 
when the gas depletion is dominated by accretion onto the BH.   
With increasing density of the ambient gas, for each $M_{\rm bh}$, 
the transition 
to mode-II oscillation and the transition to Eddington-limited regime 
happen at similar densities $n_{\rm H, \infty} = n_{\rm H, \infty}^{\rm cr}
\sim  n_{\rm H, \infty}^{\rm Edd}$. In the Eddington-limited regime
$\tau_{\rm cycle}$ becomes proportional to $\ars ^{3/2}$ for each BH
mass (dashed lines). }
\label{rs_period} \end{figure}

Figure~\ref{rs_period} shows the relationship between the period of
accretion bursts and the average size of the \str sphere produced by
the accreting BH. When the gas depletion inside the \hii~region is
dominated by the outward flow of gas toward the I-front, 
$\tau_{\rm cycle}^{\rm I}$ 
shows a linear relation with $\ars$ (solid line). This
linear relation is almost identical to the results in Paper~I, where
helium cooling/heating was not included. By increasing the ambient gas density,
eventually the gas depletion becomes dominated by accretion onto the
BH. In this latter case, assuming that the dimensionless accretion rate
$\lambdaavg$ is constant (a valid assumption in the sub-Eddington
regime), $\tau_{\rm cycle}$ scales as $\ars^3$ (dotted line). However, the
simulation results at high ambient gas density shown in
Figure~\ref{rs_period} are not well fitted by $\tau_{\rm cycle}
\propto \ars^3$, and indeed seem to follow a linear relationship
$\tau_{\rm cycle} \propto \ars$, similar to the low-density one but
with an offset. This can be explained because at high densities the
accretion rate becomes Eddington-limited soon after the transition to
mode-II accretion for which $\tau_{\rm cycle} \propto \ars^3$. It
follows that the assumption $\lambdaavg \approx {\rm const}$ becomes
invalid and instead $\tau_{\rm cycle}^{\rm II} \equiv t_{\rm in} \propto \rho
\ars^3/\dot{M}_{\rm Edd}$. In this regime, since the \str radius is
$\ars^3 \propto \eta \dot{M}_{\rm Edd}/\rho^{2}$, we get $\rho \propto
\dot{M}_{\rm Edd}^{1/2} \ars^{-1.5}$ and
\begin{equation} 
\tau_{\rm cycle}^{\rm II} \propto M_{\rm bh}^{-0.5} \ars^{1.5}.  
\end{equation}
As shown by the dashed lines in Figure~\ref{rs_period}, this model is
in good agreement with the results of the simulations for different
values of $M_{\rm bh}$.

Thus, the small offset in $\tau_{\rm cycle}$ observed in
Figure~\ref{rs_period} when the density is increased, can be
understood because $n_{\rm H, \infty}^{\rm cr}$, at which the
transition from mode-I to mode-II accretion takes place, is nearly
equal to $n_{\rm H, \infty}^{\rm Edd}$, the critical density at which the mean accretion
rate becomes Eddington limited. But, in general, the ratio of these critical
densities may depend on all the free parameters of the model. 

Our analytical model of feedback-regulated feeding of the BH, can help
understand the dependence of the critical density on all the parameter
space, not fully covered by the simulations. We found that the cycle
period $\tau_{\rm cycle}$ is the shortest time between the gas
depletion timescales $t_{\rm in} = M_{\rm \hii}/\dot{M}$, where
$M_{\rm \hii} \sim \rho_{\rm in} \ars^3$ is the mass inside the
\hii~region, and $t_{\rm out} \approx 3\ars/c_{\rm s,in}$ (see
Paper~I). Thus, by definition, when the density approaches the
critical density we have $t_{\rm in} \simeq t_{\rm out}$, but this
condition also implies that the mean \str radius approaches the
effective accretion radius:
\begin{equation}
\ars^{\rm cr} \approx 10 \times r_{\rm b,eff}.
\label{eq:rsc}
\end{equation}
Equation~(\ref{eq:rsc}), is derived setting $\dot{M} =\langle
\dot{M}\rangle \equiv 4\pi \rho_{\rm in} c_{\rm s,in}r_{\rm b,eff}^2
\equiv \lambdaavg \dot{M_{B}}$ in the relationship for
$t_{\rm in}$. Since in our model we have $\rho_{\rm in}T_{\rm in}
\simeq \rho_\infty T_\infty$, it follows that $r_{\rm b, eff} \approx
(T_{\rm in}/T_\infty)^{1/4}\lambdaavg^{1/2} r_{\rm b}$, and $\ars^{\rm
  cr} \simeq 2 T_{\infty,4} T_{\rm in,*}^{-7/4} r_{\rm b}$, where
$r_{\rm b}\equiv GM/c_{s,\infty}^2$ is the Bondi radius,
$T_{\infty,4}\equiv T_\infty/10^4$~K, and $T_{\rm in, *} \equiv T_{\rm in}/6
\times 10^4$~K is the mean temperature at the accretion radius inside
the \hii~ region (normalized to the value found for
$\alpha=1.5$). Thus, $\ars^{\rm cr}$ and period of the bursts are
\begin{eqnarray}
\ars^{\rm cr} &\approx& (0.01~{\rm pc}) \mtwo 
T_{\rm in,*}^{-7/4},\label{eq:rs}\\ 
\tau_{\rm cycle}^{\rm cr} &\approx& (1000~{\rm yr}) \mtwo T_{\rm in,*}^{-9/4},
\end{eqnarray}
with $\mtwo \equiv M_{\rm bh}/100~M_\odot$. 

Applying naively the analytical expression for the \str radius
produced by a source of luminosity $L \equiv \eta c^2 \langle
\dot{M}\rangle$ in a gas of density $\rho_\infty$ gives $\ars \propto
M^{2/3}n_{\rm H,\infty}^{-1/3}T_\infty^{1/3}\eta^{1/3}$. However, using
the simulation data, we find that the mean radius of the \str sphere
in the sub-Eddington regime is nearly independent of $T_\infty$, and
if $\ars \sim \ars^{\rm cr}$, is also independent of $\eta$:
\begin{equation}
\ars \approx (0.015~{\rm pc}) \mtwo^{2/3} 
\left( \frac{n_{\rm H,\infty}}{10^6~{\rm cm}^{-3}}\right)^{-1/3} 
\left(\frac{\bar{E}}{41~{\rm eV}}\right)^{-5/8},
\label{eq:rs_sim}
\end{equation}
where $\bar{E} \equiv L_0/S_0$ is the mean energy of ionizing photons,
and we have assumed hydrogen recombination coefficient $\alpha_R =(4
\times 10^{13}~ {\rm cm}^3/{\rm s}) T_{\rm in,*}^{-1/2}$. In addition,
we find that, $\ars \propto \eta^{1/3}$ as expected for $\ars \gg
\ars^{\rm cr}$. The deviation from the naive expectation is not
surprising, as the BH luminosity and the density inside the \str
sphere are not constant with time. Indeed, although both the maximum
and mean luminosities of the BH are $\propto \eta$, the simulations
show that the luminosity at the minimum of the cycle, $L^{\rm min}$,
is nearly independent of $\eta$. Typically $L^{\rm min}\ll L$, but
when $n_{\rm H, \infty}$ approaches the critical value $L^{\rm min} \sim
L$. Similarly, assuming an effective mean density $(\rho_{\rm
  in}\rho_\infty)^{1/2}$ in the \str radius expression would explain
the temperature dependence in Equation~(\ref{eq:rs_sim}).

Finally, setting $\ars=\ars^{\rm cr}$ we derive the critical density
\begin{equation}
n_{\rm H, \infty}^{\rm cr} \sim (5 \times 10^6~{\rm cm}^{-3}) \mtwo^{-1}
T_{\rm in,*}^{7/4} \left({\bar{E} \over 41~{\rm eV}}\right)^{-1}.
\end{equation}

\begin{figure}[t]
\epsscale{1.0} \plotone{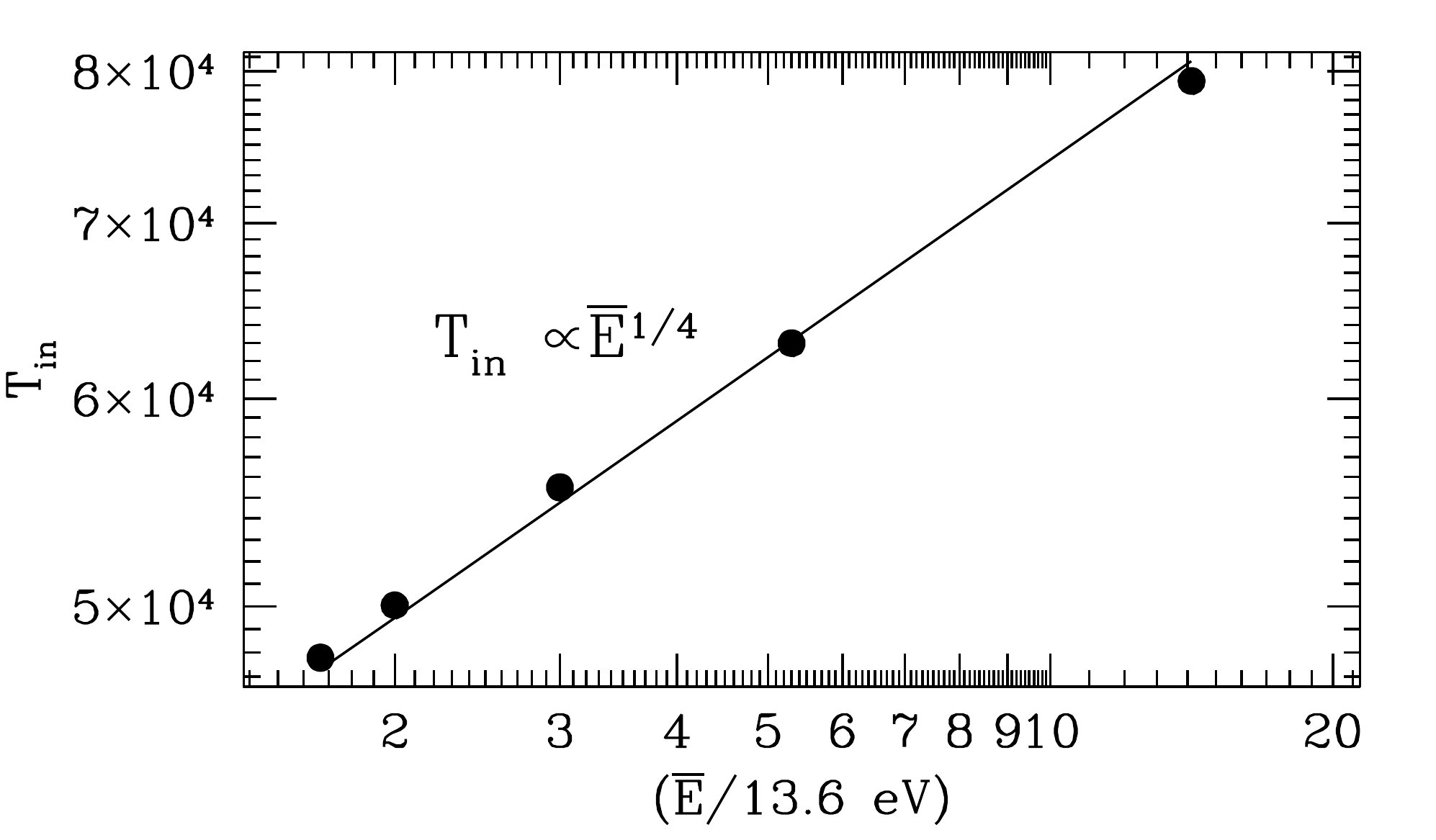}
\caption{Temperature $T_{\rm in}$ at the effective inner Bondi radius, 
located inside the \hii~region produced by the accreting BH, as a
function of the mean energy of ionizing photons $\bar{E}$ of the spectrum
of radiation emitted near the BH by the accretion disk. We have assumed
a gas of nearly zero-metallicity and a power-law spectrum $F_\nu \propto
\nu^{-\alpha}$.} 
\label{fig:E} \end{figure}

The critical density $n_{\rm H,\infty}^{\rm cr}$ as well as other
scaling relationships in our model depends on $\bar{E} \equiv L/S_0$
and $T_{\rm in}$, but for a gas of zero metallicity (including
helium), these quantities are determined only by the spectrum of the
radiation. Assuming a power law spectrum with index $\alpha$ is easy
to show that
\begin{equation}
\bar{E}=13.6~{\rm eV}
\begin{cases}
\alpha/(\alpha-1) & \mbox{if } \alpha>1,\\ 
\ln(h\nu_{\rm max}/13.6~{\rm eV}) & \mbox{if } \alpha=1\\
\alpha/(1-\alpha)(h\nu_{\rm max}/13.6~{\rm eV})^{\alpha} & \mbox{if }\alpha<1.\\
\end{cases}
\label{eq:E_alpha}
\end{equation}
We have estimated $h\nu_{\rm max}=0.2$~keV as the frequency at which the
mean free path of the photons equals $\ars$. The points in
Figure~\ref{fig:E} show $T_{\rm in}$ as a function of $\bar{E}$ for simulations
with $\alpha=0.5, 1, 1.5, 2, 2.5$ taken from Figure~9 in Paper~I. The
line shows the fit to the points:
\begin{equation}
T_{\rm in,*} \approx \left({\bar{E} \over 41~{\rm eV}}\right)^{1/4}.
\label{eq:E_ev}
\end{equation}
For our fiducial model,
for which $\bar{E} \sim 41$~eV, the value of the
critical density is very close to $n_{\rm H,\infty}^{\rm Edd}$ given in
Equation~(\ref{eq:bondi_edd}):
\begin{equation}
{n^{\rm cr}_{\rm H,\infty} \over n_{\rm H,\infty}^{\rm Edd}} \approx \eta_{-1} T_{\infty,4}
\left({\bar{E} \over 41~{\rm eV}}\right)^{-9/16}.
\label{eq:ncr}
\end{equation}
From an inspection of Equation~(\ref{eq:ncr}) is evident that the only
cases in which $n_{\rm H,\infty}^{\rm cr}$ can be larger than
$n_{\rm H,\infty}^{\rm Edd}$ are assuming the largest realistic values of unity
for $T_{\infty, 4}$ and $\eta_{-1}$, and assuming a spectrum of
radiation from the BH softer than $\alpha=1.5$ that would reduce
$\bar{E}$ with respect to the fiducial value. 
Vice versa a
hard spectrum, low radiative efficiency and accretion from a gas
colder than $10^4$~K would decrease the ratio
$n_{\rm H,\infty}^{\rm cr}/n_{\rm H,\infty}^{\rm Edd}$ below unity, making
mode-II accretion sub-Eddington for a wider range of densities. For
these cases the period of the cycle could become very short with
increasing density as $\tau_{\rm cycle} \propto \ars^3 \propto n_{\rm
H,\infty}^{-1}$.

\subsection{Effect of Non-zero Angular Momentum of Gas}\label{ssec:ang}
\label{sec:tdelay}
\begin{figure*}[t] \epsscale{1.0} \plotone{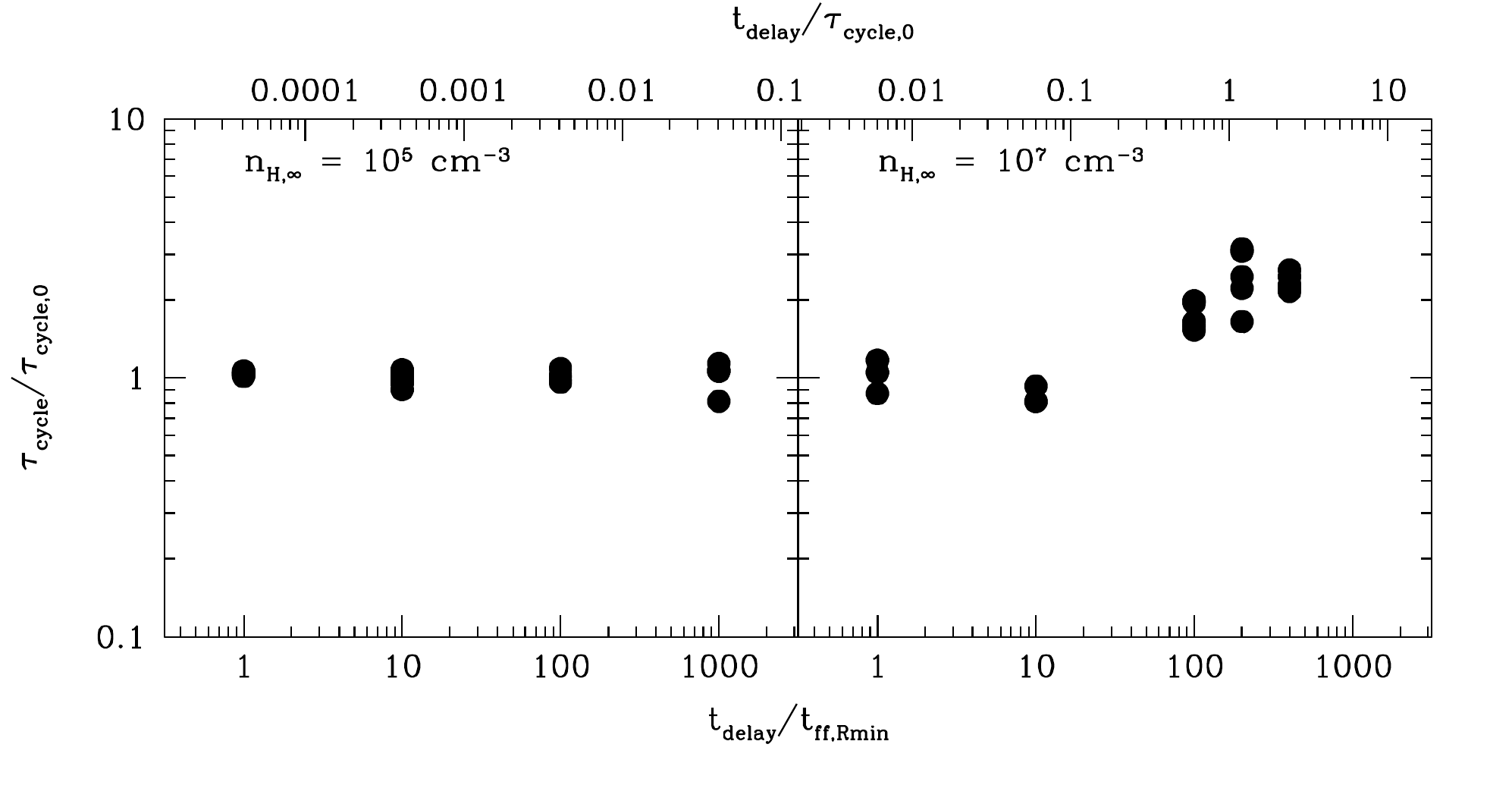} 
\caption{ Dependence of the period between bursts $\tau_{\rm cycle}$
on the time delay between the accretion rate at $R_{\rm min}$ and the
BH output luminosity. The time delay is produced by the presence of
the accretion disk that is unresolved in our simulations.  The two panels
show $\tau_{\rm cycle}$ in units of $\tau_{\rm cycle,0} \equiv \tau_{\rm
cycle}(t_{\rm delay} =0) $ as a function of $t_{\rm delay}$ for $\nfive$
(left panel) and $10^7~$cm$^{-3}$ (right panel).  The bottom axis shows
$t_{\rm delay}$ in units of the free-fall time at $R_{\rm min}$ and the
top axis in units of $\tau_{\rm cycle,0}$.  
The introduction of a time delay does
not change $\tau_{\rm cycle}$ when the gas density is $\nfive$, while
$\tau_{\rm cycle}$ increases by approximately the amount of time delay
introduced for $\nseven$. In this density regime, the largest time delays 
introduced are comparable to the oscillation period $\tau_{\rm cycle,0}$. 
In both cases, the oscillatory behavior of the accretion luminosity does not disappear. }
\label{td}
\end{figure*}

As discussed in Paper~I, the introduction of small angular momentum
in the flow, which is realistic in most astrophysical problems, can modify the
time-dependent behavior of accretion rate presented in this series
of papers. Angular momentum of gas leads to the formation of an accretion
disk near the Schwartzshild radius of a BH. This disk is not resolved in our
simulations. Thus, the accreted gas
may experience a time delay before it is converted to radiation.  Here,
we test how the introduction of time delay would affect the feedback
loops of accretion.

As mentioned in Section~2, it is important to estimate physically
motivated time delays. Here, we explore the time delay of
$1$--$300$ times $t_{\rm ff}(R_{\rm min})$ which is large enough with an
assumption of $\alpha$-disk model. On the other hand, no matter how
long is the time delay, what really matters is how the time delay
compares to the oscillation period, which depends mainly on the gas
density for a fixed mass of BH. We investigate this issue in the low
($\nfive$) and high density ($\nseven$) regimes where the oscillation
pattern and the periods are different. At low densities a time delay
of a few hundred free-fall times is much smaller compared to the
oscillation period, whereas at high densities the maximum time
delay that we have tested is comparable to the oscillation period. In
the left panel of Figure~\ref{td} which shows the result for $\nfive$,
$\tau_{\rm cycle}$ does not increase at all as a function of time
delay since the introduced time delay is much smaller than the
original oscillation period. In the right panel of Figure~\ref{td} for
$\nseven$, the maximum time delay that we introduce is comparable to
the original oscillation period, and we see that $\tau_{\rm cycle}$
increase approximately by the amount of time delay.  In both cases, we
still observe oscillations. Thus, only in the case of accretion from a
high-density gas which produces shorter oscillation period, and for an
accretion disk with $R_{\rm disk} \sim R_{\rm min}$, the time delay
may have an important effect on the accretion rate.  

Indeed, the accretion disk may not only introduce a time delay but
also smooth out the accretion rate on a timescale of the order of
the viscous timescale. In this case for cases in which the disk is
large ($R_{\rm disk} \sim R_{\rm min}$) and $\tau_{\rm cycle}$ is
short (i.e., for mode-II accretion), the disk may further smooth out or
completely erase the periodic low-amplitude oscillations in the
accretion rate from large scales.
   

\section{Summary and Discussion: Scaling Relationships}
We have presented a systematic study on how the classic Bondi problem
of spherical accretion onto a compact object is modified by the
effects of radiation feedback. 
We solve radiative transfer equations in the radial direction for
the hydrogen and helium ionizing radiation emitted by the BH. Gas
is optically thin inside Str\"{o}mgren radius while it becomes optically
thick for gas outside the ionized gas.
 In this paper, the second of a series,
we have focused on the effects that radiation pressure and angular
momentum have on the gas supply and accretion rate onto the BH. The
simulations focused on accretion onto IMBHs but the analytical scaling
relationships we have derived are rather general, and although the
initial conditions are somewhat idealized, should describe reality
more accurately than the classical Bondi formulae.

Here, we summarize the main results and scaling relationships we found in
the first two papers of this series for non-moving BHs accreting from a
uniform medium. In our models we have assumed that the BH accretes from
a uniform density and temperature reservoir, significantly larger
than the Bondi radius and $\langle R_s \rangle$. This assumption is well
motivated for accretion onto stellar and IMBH, but for SMBH there could
be supply of gas to the BH from stars within $\langle R_s \rangle$ 
(stellar winds) or
other astrophysical object (merger-driven accretion, etc).  The scaling
relationships can be applied to problems involving a wide range of masses
of the accretor, from stellar mass objects \citep[e.g.,][]{WheelerJ:11}
to supermassive BHs.  One caveat is that we are neglecting the effects
of self-gravity of the gas \citep[see][]{Li:11} and the gravitational
potential due to the dark matter halo of the host galaxy, which may play
an important role for the case of accretion onto supermassive BHs. Indeed,
a simple calculation shows that at the I-front gravity due to the mass
of the gas inside the \str sphere exceeds the BH's gravity if $M_{\rm bh}
\gtrsim 10^6~M_\odot/(\eta_{-1}T_{\infty,4})$. Our model
predicts scaling relationships for the period, duty cycle, peak and
mean accretion onto the BH, as well as relevant critical densities and
size of the \str sphere around the BH. In the following summary of the
scaling relationships, we express $T_{\rm in}$ in the equations in
terms of $\bar{E}$ given by Equation~(\ref{eq:E_ev}) that is valid for a gas
of low-metallicity. $\bar{E}$ is related to the spectral index $\alpha$ by
Equation~(\ref{eq:E_alpha}). For higher values of the gas metallicity,
the coefficients in the equations can be different due to changes in
the relationship between $T_{\rm in}$ and the spectrum of the
radiation. A caveat is that our simulations have explored a large but
limited parameter space for the masses of the BHs, temperature and
density of the ambient gas, etc. So, the proposed scaling
relationships, although they are based on a physically motivated model
we inferred from the simulations, should be used with caution for sets
of parameters that are significantly different from the range
confirmed by simulations.

The main qualitative result of our study is that radiation feedback
produces periodic oscillations of the accretion rate from large scales
onto the BH, and thus periodic short-lived bursts of the BH
luminosity. We found two modes of self-regulated accretion, determined by
\begin{equation}
n_{\rm H, \infty}^{\rm cr} \sim {5 \times 10^6~{\rm cm}^{-3} \over \mtwo}
\left({\bar{E} \over 41~{\rm eV}}\right)^{-9/16}.
\end{equation}
If $n_{\rm H,\infty}<n_{\rm H,\infty}^{\rm cr}$ (mode-I), the accretion
luminosity of the BH has regular bursts with period $\tau_{\rm cycle}^{\rm I}$
during which the BH increases its brightness by about 5 orders of
magnitude but only for a short fraction of the cycle period: the duty
cycle is $f_{\rm duty}^{\rm I} \equiv \tau_{\rm on}/\tau_{\rm cycle} \sim
6\%~T_{\infty,4}^{1/2}$.
During the quiescent phase in the accretion cycle the gas
accumulates in a dense shell in front of the \hii~region rather than
accreting directly onto the BH. As the luminosity decreases after the
burst, the density inside the \hii~ region also decreases because is
pushed outward by a pressure gradient, thus maintaining the I-front
radius nearly constant. Eventually the density and pressure inside the
\hii~region cannot sustain the weight of the dense shell that collapses
producing a burst of accretion. The cycle repeats regularly. If $n_{\rm
H,\infty}>n_{\rm H,\infty}^{\rm cr}$ (mode-II) the cycle is qualitatively
different: the duty cycle is about $f_{\rm duty}^{\rm II} \gtrsim 50\%$
and the peak accretion rate is only a few times the mean. There is no
collapse phase of the dense shell and the \hii~region remains roughly
stationary while the accretion rate oscillates. The physical
motivation for mode-II accretion is that the timescale for the
depletion of the gas inside the \hii~region becomes dominated by
accretion onto the BH. Only for mode-II accretion the BH growth rate
can approach the Eddington limit, given that the density exceeds the
critical density
\begin{equation}
n_{\rm H,\infty}^{\rm Edd} \sim {4 \times 10^6~{\rm cm}^{-3} \over \mtwo} 
T_{\infty,4}^{-1} \eta_{-1}^{-1}.
\end{equation}
For nearly all realistic cases $n_{\rm H,\infty}^{\rm Edd} \gtrsim n_{\rm
H,\infty}^{\rm cr}$. 

For $M_{\rm bh} = 100~M_{\odot}$, at densities $10^5$~cm$^{-3} \le 
n_{\rm H,\infty}\le n_{\rm H,\infty}^{\rm Edd} $ 
the {\it mean accretion rate} onto the BH, in units of
the Bondi rate is $\lambdaavg \sim 1\% T_{\infty,4}^{2.5}
(\bar{E}/{\rm 41~eV})^{-1}$, independent of all the other
parameters. For $n_{\rm H,\infty}<10^5$~cm$^{-3}$ instead, $\lambdaavg
\sim 1\% (n_{\rm H,\infty}/10^5~{\rm
  cm}^{-3})^{1/2}T_{\infty,4}^{2.5}(\bar{E}/{\rm 41~eV})^{-1}$,
depends weakly on the gas density. One caveat is that in Paper~I the
dependence on the free parameters of the transition density $\nfive$
has been only partially explored. As shown in Figure~\ref{den_mass} of the
present paper, the simulation results are consistent with a transition
density inversely proportional to the BH mass. Hence, if 
$ 10^5\mtwo^{-1}$~cm$^{-3} \le n_{\rm H,\infty} \le n_{\rm H,\infty}^{\rm Edd} $ the mean accretion rate is
proportional to the thermal pressure $n_{\rm H,\infty}T_{\infty}$ of
the ambient gas:
\begin{equation}
\langle \dot{M} \rangle \approx (4 \times10^{18}~{\rm g~s^{-1}}) \mtwo^2 
\left({n_{\rm H,\infty}} \over {10^5~{\rm cm}^{-3}}\right) T_{\rm \infty,4}
\left(\frac{\bar{E}}{\rm 41~eV}\right)^{-1}.
\end{equation}
If $n_{\rm H,\infty} > n_{\rm H,\infty}^{\rm Edd}$ then $\langle \dot{M} \rangle 
= L_{\rm Edd}(\eta c^{2})^{-1}$. The duty cycle is
\begin{equation}
f_{\rm duty} = 
\begin{cases}
f_{\rm duty}^{\rm I} \approx 6\% T_{\rm \infty, 4} ^{1/2} & \mbox{if } n_{\rm H,\infty} \le n_{\rm H,\infty}^{\rm cr}\\
f_{\rm duty}^{\rm II} \gtrsim 50\% & \mbox{if } n_{\rm H,\infty} > n_{\rm H,\infty}^{\rm cr}, 
\end{cases}
\end{equation}
and the maximum accretion luminosity which depends on the duty cycle thus is,
\begin{equation}
{L^{\rm max} \over L_{\rm Edd}} \approx \min{\left[1, {\cal A} \eta_{-1} \mtwo 
\left({n_{\rm H,\infty}} \over{10^5~{\rm cm}^{-3}}\right) T_{\rm \infty,4} 
\left(\frac{\bar{E}}{\rm 41~eV}\right)^{-1}\right]},
\end{equation}
where $L_{\rm Edd}=1.3 \times 10^{40} \mtwo~{\rm erg~s^{-1}}$, and
\begin{equation}
{\cal A} =  
\begin{cases}
{\cal A}^{\rm I} \approx 0.5~T_{\infty,4}^{-1/2} & \mbox{if } n_{\rm H,\infty} \le n_{\rm H,\infty}^{\rm cr}\\
{\cal A}^{\rm II} \approx 0.06 & \mbox{if } n_{\rm H,\infty} > n_{\rm H,\infty}^{\rm cr}\\
\end{cases}
\end{equation}
The cycle of the oscillations also falls into two regimes:
\begin{equation}
\tau_{\rm cycle} = 
\begin{cases}
\tau_{\rm cycle}^{\rm I} \approx (0.1~{\rm Myr})~\mtwo^{2/3} \eta_{-1}^{1/3}  
(\frac{n_{\rm H,\infty}}{1~{\rm cm^{-3}}})^{-1/3} (\frac{\bar{E}}{41~{\rm eV}})^{-3/4},
& \mbox{if } n_{\rm H,\infty} \le n_{\rm H,\infty}^{\rm cr}\\ 
\tau_{\rm cycle}^{\rm II} \approx (1~{\rm Gyr})~\eta_{-1} (\frac{n_{\rm H,\infty}}{1~{\rm cm}^{-3}})^{-1} 
(\frac{\bar{E}}{41~{\rm eV}})^{-7/8},&  
\mbox{if } n_{\rm H,\infty} > n_{\rm H,\infty}^{\rm cr}. \\
\end{cases}
\end{equation}

The astrophysical applications of this model are innumerable and are 
beyond the aim of this paper to discuss them in detail. However, one
of the most obvious results is that the luminosity of an accreting BH
should be smaller than the value inferred applying the Bondi
formula. Not only because the mean accretion rate is always $\lesssim
1\%$ of the Bondi rate, but also because 
if $n_{\rm H,\infty}<n_{\rm H,\infty}^{\rm cr}$, 94\% of the
time is about 5 orders of magnitude lower than the Bondi rate inferred
from the ambient medium temperature and density.  Thus, this simple
arguments could have interesting consequences to interpret the observed
quiescence of SMBH in ellipticals and Sgr~A$^*$. Also, the duty cycle
of $\sim 6\%~T_{\infty,4}^{1/2}$ 
we found for mode-I accretion is interestingly close to
the fraction of galaxies with AGNs $\sim 3\%$ found deep
field surveys \cite[e.g.,][]{Steideletal:02,Luoetal:11}. For SMBHs of
about $10^6$~$M_\odot$, $n_{\rm H,\infty}^{\rm cr} 
\sim 500~{\rm cm}^{-3}$.
 
In this paper we also found that IMBH can grow at near the Eddington
limit if $n_{\rm H,\infty}> {\rm max}(n_{\rm H,\infty}^{\rm Edd},n_{\rm
H,\infty}^{\rm cr})$.  This has potentially important consequences on
the ability of seed IMBH from Population~III stars to grow by accretion into
SMBH during the first gigayears of the universe age. This possibility seemed
precluded if the duty cycle of the burst was $6\%$ as found in
previous works. 

Finally, although the nature of ULXs is unknown, there are indications
that they may host an IMBH \citep[e.g.,][]{StrohmayerM:09}.  An IMBH
accreting from an interstellar medium (ISM) with high pressure such as dense molecular cloud 
($n_{\rm H,\infty} T_{\infty} \sim 
10^5$--$10^7~{\rm cm}^{-3}~{\rm K}$) 
would be $L^{\rm max} \sim 10^{37}$--$10^{39}~{\rm erg~s^{-1}}$ for 
$M_{\rm bh}=1000~M_\odot$, that is comparable to the luminosity of
ULXs. However, this assumes that the IMBH is at rest with respect to
the ISM. We will focus on gas accretion onto moving BHs with radiation
feedback in the third paper of this series (K. Park \& M. Ricotti 2012,
in preparation). Clearly, more work is needed to address each of the
aforementioned topics in detail, but the basic ground work presented
in the present paper may allow the re-visitation of a few longstanding
problems still unsolved in astrophysics.

\acknowledgments
The authors thank the referee, Luca Ciotti, Richard Mushotzky, Chris Reynolds, Eve Ostriker, and
Edward Shaya for constructive comments and feedback. The simulations
presented in this paper were carried out using high-performance computing
clusters administered by the Center for Theory and Computation of
the Department of Astronomy at the University of Maryland ("yorp"),
and the Office of Information Technology at the University of Maryland
("deepthought"). This research was supported by NASA grants NNX07AH10G
and NNX10AH10G.

\bibliographystyle{apj}
\bibliography{pr11_v2}

\begin{thebibliography}{68}
\expandafter\ifx\csname natexlab\endcsname\relax\def\natexlab#1{#1}\fi

\bibitem[{{Abel} {et~al.}(1998){Abel}, {Anninos}, {Norman}, \&
  {Zhang}}]{AbelANZ:98}
{Abel}, T., {Anninos}, P., {Norman}, M.~L., \& {Zhang}, Y. 1998, \apj, 508, 518

\bibitem[{{Abel} {et~al.}(2000){Abel}, {Bryan}, \& {Norman}}]{AbelBN:00}
{Abel}, T., {Bryan}, G.~L., \& {Norman}, M.~L. 2000, \apj, 540, 39

\bibitem[{{Alvarez} {et~al.}(2009){Alvarez}, {Wise}, \& {Abel}}]{AlvarezWA:09}
{Alvarez}, M.~A., {Wise}, J.~H., \& {Abel}, T. 2009, \apjl, 701, L133

\bibitem[{{Begelman}(1985)}]{Begelman:85}
{Begelman}, M.~C. 1985, \apj, 297, 492

\bibitem[{{Begelman} {et~al.}(2006){Begelman}, {Volonteri}, \&
  {Rees}}]{BegelmanVR:06}
{Begelman}, M.~C., {Volonteri}, M., \& {Rees}, M.~J. 2006, \mnras, 370, 289

\bibitem[{{Bisnovatyi-Kogan} \& {Blinnikov}(1980)}]{BB:80}
{Bisnovatyi-Kogan}, G.~S., \& {Blinnikov}, S.~I. 1980, \mnras, 191, 711

\bibitem[{{Bondi}(1952)}]{Bondi:52}
{Bondi}, H. 1952, \mnras, 112, 195

\bibitem[{{Bondi} \& {Hoyle}(1944)}]{BondiH:44}
{Bondi}, H., \& {Hoyle}, F. 1944, \mnras, 104, 273

\bibitem[{{Bromm} {et~al.}(1999){Bromm}, {Coppi}, \& {Larson}}]{BrommCL:99}
{Bromm}, V., {Coppi}, P.~S., \& {Larson}, R.~B. 1999, \apjl, 527, L5

\bibitem[{{Carr} {et~al.}(1984){Carr}, {Bond}, \& {Arnett}}]{Carr:84}
{Carr}, B.~J., {Bond}, J.~R., \& {Arnett}, W.~D. 1984, \apj, 277, 445

\bibitem[{{Ciotti} \& {Ostriker}(2007)}]{CiottiO:07}
{Ciotti}, L., \& {Ostriker}, J.~P. 2007, \apj, 665, 1038

\bibitem[{{Ciotti} {et~al.}(2009){Ciotti}, {Ostriker}, \&
  {Proga}}]{CiottiOP:09}
{Ciotti}, L., {Ostriker}, J.~P., \& {Proga}, D. 2009, \apj, 699, 89

\bibitem[{{Cowie} {et~al.}(1978){Cowie}, {Ostriker}, \& {Stark}}]{CowieOS:78}
{Cowie}, L.~L., {Ostriker}, J.~P., \& {Stark}, A.~A. 1978, \apj, 226, 1041

\bibitem[{{Di Matteo} {et~al.}(2008){Di Matteo}, {Colberg}, {Springel},
  {Hernquist}, \& {Sijacki}}]{DiMatteo:08}
{Di Matteo}, T., {Colberg}, J., {Springel}, V., {Hernquist}, L., \& {Sijacki},
  D. 2008, \apj, 676, 33

\bibitem[{{Fryer} {et~al.}(2001){Fryer}, {Woosley}, \& {Heger}}]{Fryer:01}
{Fryer}, C.~L., {Woosley}, S.~E., \& {Heger}, A. 2001, \apj, 550, 372

\bibitem[{{Greif} {et~al.}(2008){Greif}, {Johnson}, {Klessen}, \&
  {Bromm}}]{Greif:08}
{Greif}, T.~H., {Johnson}, J.~L., {Klessen}, R.~S., \& {Bromm}, V. 2008,
  \mnras, 387, 1021

\bibitem[{{Haehnelt} {et~al.}(1998){Haehnelt}, {Natarajan}, \&
  {Rees}}]{HaehneltNR:98}
{Haehnelt}, M.~G., {Natarajan}, P., \& {Rees}, M.~J. 1998, \mnras, 300, 817

\bibitem[{{Hayes} {et~al.}(2006){Hayes}, {Norman}, {Fiedler}, {Bordner}, {Li},
  {Clark}, {ud-Doula}, \& {Mac Low}}]{Hayes:06}
{Hayes}, J.~C., {Norman}, M.~L., {Fiedler}, R.~A., {Bordner}, J.~O., {Li},
  P.~S., {Clark}, S.~E., {ud-Doula}, A., \& {Mac Low}, M.-M. 2006, \apjs, 165,
  188

\bibitem[{{Johnson} \& {Bromm}(2007)}]{JohnsonB:07}
{Johnson}, J.~L., \& {Bromm}, V. 2007, \mnras, 374, 1557

\bibitem[{{Johnson} {et~al.}(2011){Johnson}, {Khochfar}, {Greif}, \&
  {Durier}}]{JohnsonKGD:11}
{Johnson}, J.~L., {Khochfar}, S., {Greif}, T.~H., \& {Durier}, F. 2011, \mnras,
  410, 919

\bibitem[{{Kim} {et~al.}(2011){Kim}, {Wise}, {Alvarez}, \& {Abel}}]{KimWAA:11}
{Kim}, J.-h., {Wise}, J.~H., {Alvarez}, M.~A., \& {Abel}, T. 2011, \apj, 738,
  54

\bibitem[{{Krolik}(2004)}]{Krolik:04}
{Krolik}, J.~H. 2004, \apj, 615, 383

\bibitem[{{Krolik} \& {Kallman}(1984)}]{Krolik:84}
{Krolik}, J.~H., \& {Kallman}, T.~R. 1984, \apj, 286, 366

\bibitem[{{Krolik} \& {London}(1983)}]{KrolikL:83}
{Krolik}, J.~H., \& {London}, R.~A. 1983, \apj, 267, 18

\bibitem[{{Krolik} {et~al.}(1981){Krolik}, {McKee}, \& {Tarter}}]{Krolik:81}
{Krolik}, J.~H., {McKee}, C.~F., \& {Tarter}, C.~B. 1981, \apj, 249, 422

\bibitem[{{Kurosawa} \& {Proga}(2009{\natexlab{a}})}]{KurosawaP:09a}
{Kurosawa}, R., \& {Proga}, D. 2009{\natexlab{a}}, \mnras, 397, 1791

\bibitem[{{Kurosawa} \& {Proga}(2009{\natexlab{b}})}]{KurosawaP:09b}
---. 2009{\natexlab{b}}, \apj, 693, 1929

\bibitem[{{Kurosawa} {et~al.}(2009){Kurosawa}, {Proga}, \&
  {Nagamine}}]{KurosawaPN:09}
{Kurosawa}, R., {Proga}, D., \& {Nagamine}, K. 2009, \apj, 707, 823

\bibitem[{{Li}(2011)}]{Li:11}
{Li}, Y. 2011, ArXiv e-prints

\bibitem[{{Luo} {et~al.}(2011){Luo}, {Brandt}, {Xue}, {Alexander}, {Brusa},
  {Bauer}, {Comastri}, {Fabian}, {Gilli}, {Lehmer}, {Rafferty}, {Schneider}, \&
  {Vignali}}]{Luoetal:11}
{Luo}, B., {et~al.} 2011, \apj, 740, 37

\bibitem[{{Lusso} \& {Ciotti}(2011)}]{LussoC:11}
{Lusso}, E., \& {Ciotti}, L. 2011, \aap, 525, A115

\bibitem[{{Mack} {et~al.}(2007){Mack}, {Ostriker}, \& {Ricotti}}]{MackOR:07}
{Mack}, K.~J., {Ostriker}, J.~P., \& {Ricotti}, M. 2007, \apj, 665, 1277

\bibitem[{{Madau} \& {Rees}(2001)}]{MadauR:01}
{Madau}, P., \& {Rees}, M.~J. 2001, \apjl, 551, L27

\bibitem[{{Mayer} {et~al.}(2010){Mayer}, {Kazantzidis}, {Escala}, \&
  {Callegari}}]{MayerKEC:10}
{Mayer}, L., {Kazantzidis}, S., {Escala}, A., \& {Callegari}, S. 2010, \nat,
  466, 1082

\bibitem[{{Miller} \& {Colbert}(2004)}]{Miller:04}
{Miller}, M.~C., \& {Colbert}, E.~J.~M. 2004, International Journal of Modern
  Physics D, 13, 1

\bibitem[{{Milosavljevi{\'c}} {et~al.}(2009{\natexlab{a}}){Milosavljevi{\'c}},
  {Bromm}, {Couch}, \& {Oh}}]{MiloBCO:09}
{Milosavljevi{\'c}}, M., {Bromm}, V., {Couch}, S.~M., \& {Oh}, S.~P.
  2009{\natexlab{a}}, \apj, 698, 766

\bibitem[{{Milosavljevi{\'c}} {et~al.}(2009{\natexlab{b}}){Milosavljevi{\'c}},
  {Couch}, \& {Bromm}}]{MiloCB:09}
{Milosavljevi{\'c}}, M., {Couch}, S.~M., \& {Bromm}, V. 2009{\natexlab{b}},
  \apjl, 696, L146

\bibitem[{{Novak} {et~al.}(2011){Novak}, {Ostriker}, \& {Ciotti}}]{NovakOC:11}
{Novak}, G.~S., {Ostriker}, J.~P., \& {Ciotti}, L. 2011, \apj, 737, 26

\bibitem[{{Oh} \& {Haiman}(2002)}]{OhH:02}
{Oh}, S.~P., \& {Haiman}, Z. 2002, \apj, 569, 558

\bibitem[{{Omukai} {et~al.}(2008){Omukai}, {Schneider}, \&
  {Haiman}}]{OmukaiSH:08}
{Omukai}, K., {Schneider}, R., \& {Haiman}, Z. 2008, \apj, 686, 801

\bibitem[{{Ostriker} {et~al.}(2010){Ostriker}, {Choi}, {Ciotti}, {Novak}, \&
  {Proga}}]{OstrikerCCNP:10}
{Ostriker}, J.~P., {Choi}, E., {Ciotti}, L., {Novak}, G.~S., \& {Proga}, D.
  2010, \apj, 722, 642

\bibitem[{{Ostriker} {et~al.}(1976){Ostriker}, {Weaver}, {Yahil}, \&
  {McCray}}]{OstrikerWYM:76}
{Ostriker}, J.~P., {Weaver}, R., {Yahil}, A., \& {McCray}, R. 1976, \apjl, 208,
  L61

\bibitem[{{Park} \& {Ricotti}(2011)}]{ParkR:11}
{Park}, K., \& {Ricotti}, M. 2011, \apj, 739, 2

\bibitem[{{Pelupessy} {et~al.}(2007){Pelupessy}, {Di Matteo}, \&
  {Ciardi}}]{Pelupessy:07}
{Pelupessy}, F.~I., {Di Matteo}, T., \& {Ciardi}, B. 2007, \apj, 665, 107

\bibitem[{{Proga}(2007)}]{Proga:07}
{Proga}, D. 2007, \apj, 661, 693

\bibitem[{{Proga} {et~al.}(2008){Proga}, {Ostriker}, \&
  {Kurosawa}}]{ProgaOK:08}
{Proga}, D., {Ostriker}, J.~P., \& {Kurosawa}, R. 2008, \apj, 676, 101

\bibitem[{{Regan} \& {Haehnelt}(2009)}]{ReganH:09}
{Regan}, J.~A., \& {Haehnelt}, M.~G. 2009, \mnras, 396, 343

\bibitem[{{Ricotti}(2007)}]{Ricotti:07}
{Ricotti}, M. 2007, \apj, 662, 53

\bibitem[{{Ricotti}(2009)}]{Ricotti:09}
---. 2009, \mnras, 392, L45

\bibitem[{{Ricotti} {et~al.}(2001){Ricotti}, {Gnedin}, \&
  {Shull}}]{RicottiGS:01}
{Ricotti}, M., {Gnedin}, N.~Y., \& {Shull}, J.~M. 2001, \apj, 560, 580

\bibitem[{{Ricotti} \& {Ostriker}(2004)}]{RicottiO:04b}
{Ricotti}, M., \& {Ostriker}, J.~P. 2004, \mnras, 352, 547

\bibitem[{{Ricotti} {et~al.}(2005){Ricotti}, {Ostriker}, \&
  {Gnedin}}]{RicottiOG:05}
{Ricotti}, M., {Ostriker}, J.~P., \& {Gnedin}, N.~Y. 2005, \mnras, 357, 207

\bibitem[{{Ricotti} {et~al.}(2008){Ricotti}, {Ostriker}, \&
  {Mack}}]{Ricotti:08}
{Ricotti}, M., {Ostriker}, J.~P., \& {Mack}, K.~J. 2008, \apj, 680, 829

\bibitem[{{Sazonov} {et~al.}(2005){Sazonov}, {Ostriker}, {Ciotti}, \&
  {Sunyaev}}]{Sazonov:05}
{Sazonov}, S.~Y., {Ostriker}, J.~P., {Ciotti}, L., \& {Sunyaev}, R.~A. 2005,
  \mnras, 358, 168

\bibitem[{{Schneider} {et~al.}(2002){Schneider}, {Ferrara}, {Natarajan}, \&
  {Omukai}}]{SchneiderFNO:02}
{Schneider}, R., {Ferrara}, A., {Natarajan}, P., \& {Omukai}, K. 2002, \apj,
  571, 30

\bibitem[{{Shakura} \& {Sunyaev}(1973)}]{ShakuraS:1973}
{Shakura}, N.~I., \& {Sunyaev}, R.~A. 1973, \aap, 24, 337

\bibitem[{{Shapiro}(1973)}]{Shapiro:73}
{Shapiro}, S.~L. 1973, \apj, 180, 531

\bibitem[{{Steidel} {et~al.}(2002){Steidel}, {Hunt}, {Shapley}, {Adelberger},
  {Pettini}, {Dickinson}, \& {Giavalisco}}]{Steideletal:02}
{Steidel}, C.~C., {Hunt}, M.~P., {Shapley}, A.~E., {Adelberger}, K.~L.,
  {Pettini}, M., {Dickinson}, M., \& {Giavalisco}, M. 2002, \apj, 576, 653

\bibitem[{{Stone} \& {Norman}(1992)}]{StoneN:92}
{Stone}, J.~M., \& {Norman}, M.~L. 1992, \apjs, 80, 753

\bibitem[{{Strohmayer} \& {Mushotzky}(2009)}]{StrohmayerM:09}
{Strohmayer}, T.~E., \& {Mushotzky}, R.~F. 2009, \apj, 703, 1386

\bibitem[{{van der Marel}(2004)}]{vanderMarel:04}
{van der Marel}, R.~P. 2004, Coevolution of Black Holes and Galaxies, 37

\bibitem[{{Vitello}(1984)}]{Vitello:84}
{Vitello}, P. 1984, \apj, 284, 394

\bibitem[{{Volonteri} {et~al.}(2003){Volonteri}, {Haardt}, \&
  {Madau}}]{VolonteriHM:03}
{Volonteri}, M., {Haardt}, F., \& {Madau}, P. 2003, \apj, 582, 559

\bibitem[{{Volonteri} {et~al.}(2008){Volonteri}, {Lodato}, \&
  {Natarajan}}]{VolonteriLN:08}
{Volonteri}, M., {Lodato}, G., \& {Natarajan}, P. 2008, \mnras, 383, 1079

\bibitem[{{Volonteri} \& {Rees}(2005)}]{Volonteri:05}
{Volonteri}, M., \& {Rees}, M.~J. 2005, \apj, 633, 624

\bibitem[{{Wandel} {et~al.}(1984){Wandel}, {Yahil}, \& {Milgrom}}]{WandelYM:84}
{Wandel}, A., {Yahil}, A., \& {Milgrom}, M. 1984, \apj, 282, 53

\bibitem[{{Wheeler} \& {Johnson}(2011)}]{WheelerJ:11}
{Wheeler}, J.~C., \& {Johnson}, V. 2011, \apj, 738, 163

\bibitem[{{Yoo} \& {Miralda-Escud{\'e}}(2004)}]{YooM:04}
{Yoo}, J., \& {Miralda-Escud{\'e}}, J. 2004, \apjl, 614, L25

\end{thebibliography}



\end{document}